\documentclass[10pt, conference, compsocconf]{IEEEtran}
\usepackage{mathptmx} % This is Times font

\usepackage{fancyhdr}
\usepackage[normalem]{ulem}
\usepackage[hyphens]{url}
\usepackage[sort,nocompress]{cite}
\usepackage[final]{microtype}
\usepackage[bookmarks=true,breaklinks=true,letterpaper=true,colorlinks,citecolor=blue,linkcolor=blue,urlcolor=blue]{hyperref}

\usepackage{amsmath,amssymb,amsfonts}
\usepackage{algorithm, algpseudocode}
\usepackage{algorithmicx}
\usepackage{graphicx}
\usepackage{textcomp}
\usepackage{xcolor}
\usepackage{booktabs}

\let\labelindent\relax % To resolve enumitem error
\usepackage{enumitem}

\usepackage{tikz}

\ifCLASSINFOpdf
\else
\fi
\usepackage{xspace} % To resolve sysname error
\newcommand{\sysname}{DFX\xspace}

\makeatletter
\newcommand{\newlineauthors}{%
  \end{@IEEEauthorhalign}\hfill\mbox{}\par
  \mbox{}\hfill\begin{@IEEEauthorhalign}
}
\makeatother

\begin{document}
%
% paper title
% can use linebreaks \\ within to get better formatting as desired
\title{\sysname: A Low-latency Multi-FPGA Appliance for Accelerating Transformer-based Text Generation}

% author names and affiliations
% use a multiple column layout for up to two different
% affiliations
\author
{
\IEEEauthorblockN{\large Seongmin Hong}
\IEEEauthorblockA{\textnormal{seongminhong@kaist.ac.kr}\\
\textnormal{KAIST}\\\textnormal{Daejeon, South Korea}}
\and
\IEEEauthorblockN{\large Seungjae Moon}
\IEEEauthorblockA{\textnormal{sjaemoon@kaist.ac.kr}\\
\textnormal{KAIST}\\\textnormal{Daejeon, South Korea}}
\and
\IEEEauthorblockN{\large Junsoo Kim}
\IEEEauthorblockA{\textnormal{junsoo999@kaist.ac.kr}\\
\textnormal{KAIST}\\\textnormal{Daejeon, South Korea}}
\newlineauthors
\IEEEauthorblockN{\large Sungjae Lee}
\IEEEauthorblockA{\textnormal{sung-jae.lee@navercorp.com}\\
\textnormal{NAVER CLOVA}\\\textnormal{Seongnam, South Korea}}
\and
\IEEEauthorblockN{\large Minsub Kim}
\IEEEauthorblockA{\textnormal{minssub.kim@navercorp.com}\\
\textnormal{NAVER CLOVA}\\\textnormal{Seongnam, South Korea}}
\and
\IEEEauthorblockN{\large Dongsoo Lee}
\IEEEauthorblockA{\textnormal{dongsoo.lee@navercorp.com}\\
\textnormal{NAVER CLOVA}\\\textnormal{Seongnam, South Korea}}
% \newlineauthors
\and
\IEEEauthorblockN{\large Joo-Young Kim}
\IEEEauthorblockA{\textnormal{jooyoung1203@kaist.ac.kr}\\
\textnormal{KAIST}\\\textnormal{Daejeon, South Korea}}
}
% conference papers do not typically use \thanks and this command
% is locked out in conference mode. If really needed, such as for
% the acknowledgment of grants, issue a \IEEEoverridecommandlockouts
% after \documentclass

% for over three affiliations, or if they all won't fit within the width
% of the page, use this alternative format:
% 
% \author{\IEEEauthorblockN{Seongmin Hong\IEEEauthorrefmark{1},
% Seungjae Moon\IEEEauthorrefmark{1},
% Junsoo Kim\IEEEauthorrefmark{1}, 
% Sungjae Lee\IEEEauthorrefmark{2}, 
% Minsub Kim\IEEEauthorrefmark{2}, 
% Dongsoo Lee\IEEEauthorrefmark{2} and
% Joo-Young Kim\IEEEauthorrefmark{1}}
% \IEEEauthorblockA{Korea Advanced Institute of Science and Technology (KAIST)\IEEEauthorrefmark{1}, NAVER CLOVA\IEEEauthorrefmark{2}}
% \IEEEauthorblockA{\texttt{\{seongminhong, sjaemoon, junsoo999, jooyoung1203\}@kaist.ac.kr}\IEEEauthorrefmark{1}}
% \IEEEauthorblockA{\texttt{\{sung-jae.lee, minssub.kim, dongsoo.lee\}@navercorp.com}\IEEEauthorrefmark{2}}
% }

% use for special paper notices
%\IEEEspecialpapernotice{(Invited Paper)}

% make the title area
\maketitle

\begin{abstract}

% abstract

% - GPT is widely used in datacenter

% - GPT size is very increased, so it is difficult to compute with conventional system. therefore datacenter need accelerator for GPT

% - we proposed accelerator 1) high speed, low latency, 2) all GPT layers, 3) Parallelism and Scalability

% - we implement multi-FPGA based acceleration with model parallelism

% - We show X.0 \textendash{} XX.0$\times$ speedup over Nvidia V100 GPU

% Transformer is a deep learning language model widely used for natural language processing (NLP) applications in datacenters. Transformer models that target text generation such as Generative Pre-trained Transformer 2 (GPT-2) require efficient processing of large input context in the summarization stage followed by sequential single input in the generation stage. The conventional platforms such as GPU are specialized for the parallel processing of large inputs in the summarization stage, but their performance significantly degrades in the generation stage due to its sequential characteristic. Moreover, the datacenter requires a scalable hardware platform to handle the exponential increase of transformer model size, but scaling out these devices is expensive due to the high upfront cost and operating cost. 

Transformer is a deep learning language model widely used for natural language processing (NLP) services in datacenters. Among transformer models, Generative Pre-trained Transformer (GPT) has achieved remarkable performance in text generation, or natural language generation (NLG), which needs the processing of a large input context in the summarization stage, followed by the generation stage that produces a single word at a time. The conventional platforms such as GPU are specialized for the parallel processing of large inputs in the summarization stage, but their performance significantly degrades in the generation stage due to its sequential characteristic. Therefore, an efficient hardware platform is required to address the high latency caused by the sequential characteristic of text generation.

In this paper, we present \sysname, a multi-FPGA acceleration appliance that executes GPT-2 model inference end-to-end with low latency and high throughput in both summarization and generation stages. \sysname{} uses model parallelism and optimized dataflow that is model-and-hardware-aware for fast simultaneous workload execution among devices. Its compute cores operate on custom instructions and provide GPT-2 operations end-to-end. We implement the proposed hardware architecture on four Xilinx Alveo U280 FPGAs and utilize all of the channels of the high bandwidth memory (HBM) and the maximum number of compute resources for high hardware efficiency. \sysname{} achieves 5.58$\times$ speedup and 3.99$\times$ energy efficiency over four NVIDIA V100 GPUs on the modern GPT-2 model. \sysname{} is also 8.21$\times$ more cost-effective than the GPU appliance, suggesting that it is a promising solution for text generation workloads in cloud datacenters.

\end{abstract}

% \vspace{-0.05in}


% For peer review papers, you can put extra information on the cover
% page as needed:
% \ifCLASSOPTIONpeerreview
% \begin{center} \bfseries EDICS Category: 3-BBND \end{center}
% \fi
%
% For peerreview papers, this IEEEtran command inserts a page break and
% creates the second title. It will be ignored for other modes.
\IEEEpeerreviewmaketitle

%%%%%% -- PAPER CONTENT STARTS-- %%%%%%%%

% \vspace{0.1in}

\section{Introduction}
\label{introduction}
Transformer\cite{vaswani2017attention} is a deep learning language model that uses the mechanism of attention, which gives a different weight of significance to each part of the input data. By solving the recursion and lack of global dependency problem of recurrent neural network (RNN)\cite{cho2014learning} and long short-term memory (LSTM)\cite{hochreiter1997long}, the transformer is becoming the de facto standard for natural language processing (NLP) applications such as text generation\cite{zhang2020pointer, pascual2021plug}, text classification\cite{gonzalez2020comparing, garg2020bae}, and machine translation\cite{wu2016google, wang2019learning}. Among them, text generation, broadly referred to as natural language generation (NLG), is related to the automatic generation of human-readable text by a computer. It has become of great importance in emerging applications such as dialogue system~\cite{lin2020caire, budzianowski2019hello, wolf2019transfertransfo} and topic-to-essay generation~\cite{feng2018toe, yang2019toe, qiao2020toe}, with a rapid growth rate of 20\%~\cite{nlg} in the NLG market.  
Among transformer models, the Generative Pre-trained Transformer (GPT) is widely used in cloud services, achieving remarkable performance particularly in text generation applications. 

%%%%%%%%%%%%%%%%%%%%%%%%%%%%%%%%%%%%%%%%%%%%%%%%%%
% Figures
%%%%%%%%%%%%%%%%%%%%%%%%%%%%%%%%%%%%%%%%%%%%%%%%%%
\begin{figure}[t] 
% \vspace{-0.01in}
\centering
\footnotesize
\includegraphics[width=3.33in]{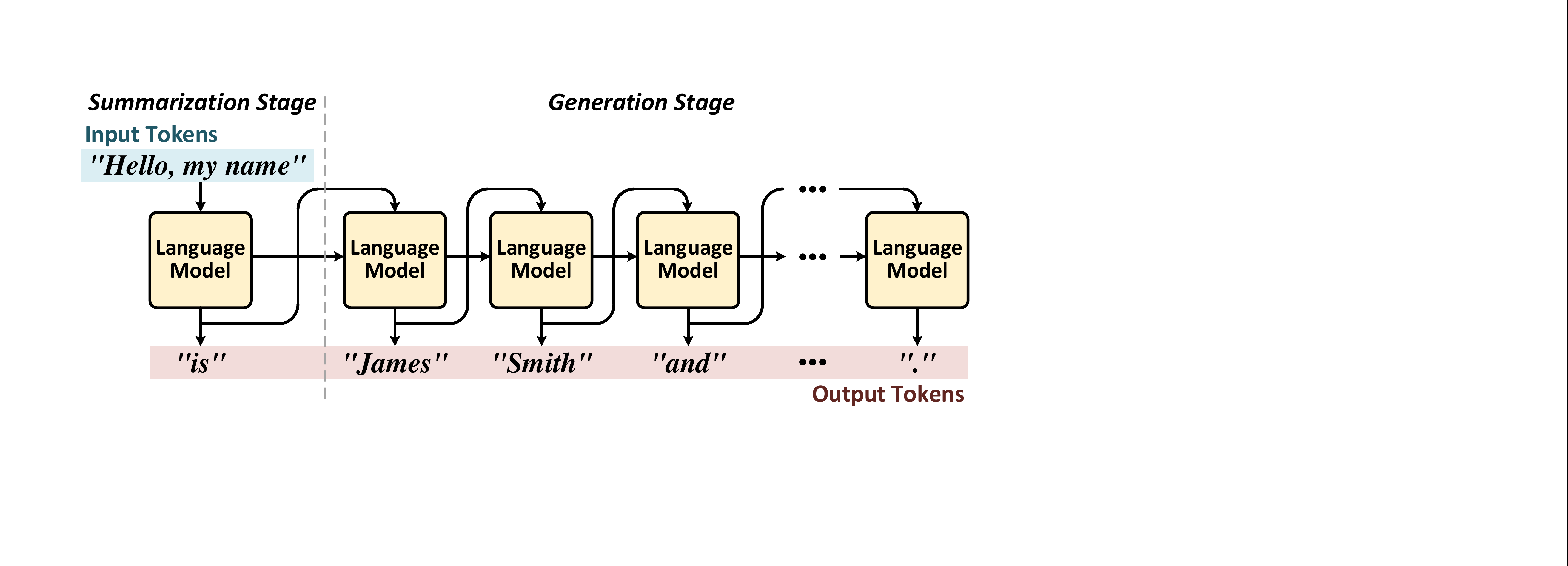} 
\vspace{-0.1in}
\caption{Illustration of transformer-based text generation.}
\label{fig-intro} 
\vspace{0.05in}
\end{figure}
%%%%%%%%%%%%%%%%%%%%%%%%%%%%%%%%%%%%%%%%%%%%%%%%%%

In the text generation process, consisting of the summarization and generation stages, the language model continuously generates sequential output words (i.e., output tokens) using the input context made of multiple input words (i.e., input tokens), as shown in Figure \ref{fig-intro}. 
%Text generation consists of summarization and generation stage.
In the summarization stage, the language model processes a batch of input tokens with a single run and generates a new output token. The generation stage iterates the language model processing to generate the subsequent output tokens, in which each iteration takes the single output token from the previous iteration as input to generate a single output token. Meanwhile, the language model accumulates the contextual features throughout the iterations.
%The single word is processed with the accumulation of intermediate features from previous language models in each iteration to retain the contextual information. 
In current server platforms, GPU\cite{dgx} is used to accelerate text generation. Its massively parallel compute units yield high performance in the summarization stage as the input tokens can be computed simultaneously. However, a significant performance degradation occurs in the generation stage because GPU is not suitable for sequential processing, suffering from severe underutilization.

% However, they are not applicable for transformer-based language services because they only accelerate a portion of the transformer process. In particular, the attention process has been their primary operation of concern because it is mostly composed of matrix multiplications that are computationally intensive but straightforward to accelerate. Moreover, the architectures of the previous works have not considered for the end-to-end acceleration of GPT because they use dedicated compute components for the few operations. 

Several architectures~\cite{ham2020a3, ham2021elsa, wang2021spatten, lu2021sanger} have been proposed to accelerate the transformer. The attention mechanism \cite{vaswani2017attention}, composed of matrix multiplication and softmax for contextual understanding, has been their primary operation of concern because it is the most computationally intensive task in the transformer. However, a language service requires an architecture that considers the entirety of the transformer model. For datacenters to adopt the above accelerator architectures, the server platforms would need CPU or extra compute modules to cover the complete operations, which would lead to large processing overhead. Therefore, a unified and programmable architecture that can support the whole GPT operations end-to-end is necessary.
%an architecture that supports GPT operations end-to-end and contains instruction-based programmable compute cores is necessary. 

In this paper, we propose \sysname, a multi-FPGA acceleration appliance that specializes in text generation workloads covering end-to-end inference of variously sized GPT models. To address the sequential characteristic of text generation, \sysname{} compute core is optimized for single token processing, which is impracticable in GPU. It also uses an efficient tiling scheme and dataflow based on the characteristics of GPT for maximum high bandwidth memory (HBM) \cite{lee201425} bandwidth usage. To address the increasing model size, \sysname{} uses model parallelism on the multi-device system to increase the physical number of compute cores that work in parallel while evenly assigning full workload to each device. Furthermore, we exploit FPGAs because the transformer-based model continues to undergo modifications and expansions for different language services in the datacenter. The FPGA-based accelerator provides fully reprogrammable hardware to support new operations and larger dimensions of the evolving transformer with minimum cost for redesign when compared to an ASIC-based accelerator.

The main contributions of our work are as follows.

\begin{itemize}[leftmargin=*]
\setlength\itemsep{0.01in}
% \vspace{-0.07in}
\item {We identify that the generation stage of text generation workload is the bottleneck on parallel hardware such as GPU due to its sequential characteristic.}
\item {We design a custom programmable compute core optimized for the end-to-end acceleration of GPT inference with a high hardware utilization.}
\item {We utilize the full HBM bandwidth complemented with an efficient tiling scheme and dataflow based on the characteristics of GPT to achieve low latency and high throughput.}
\item{We apply model parallelism and efficient network to the multi-FPGA system by evenly distributing the model parameters to each FPGA in a way that requires minimal data synchronization among FPGAs and achieves maximal parallel computation.}
% \item {We design the multi-FPGA system based on a lightweight router enabling FPGA-to-FPGA communication and apply model parallelism by evenly distributing the model parameters to each FPGA in a way that requires minimal data synchronization among FPGAs and achieves maximal parallel computation.}
\item {We build a multi-FPGA appliance with low upfront and operating cost that accelerates transformer-based language services, achieving multiple times better performance and efficiency than the conventional GPU platform. We believe this new hardware platform is promising for handling ever-increasing text generation workloads in datacenters.}
\end{itemize}

% \vspace{0.1in}

% \vspace{-0.2in}

\section{Background}
\label{background}

% \vspace{-0.05in}
\subsection{GPT Language Model} \label{background_gpt}

\begin{figure}[t] 
% \vspace{-0.01in}
\centering
\footnotesize
\includegraphics[width=3.33in]{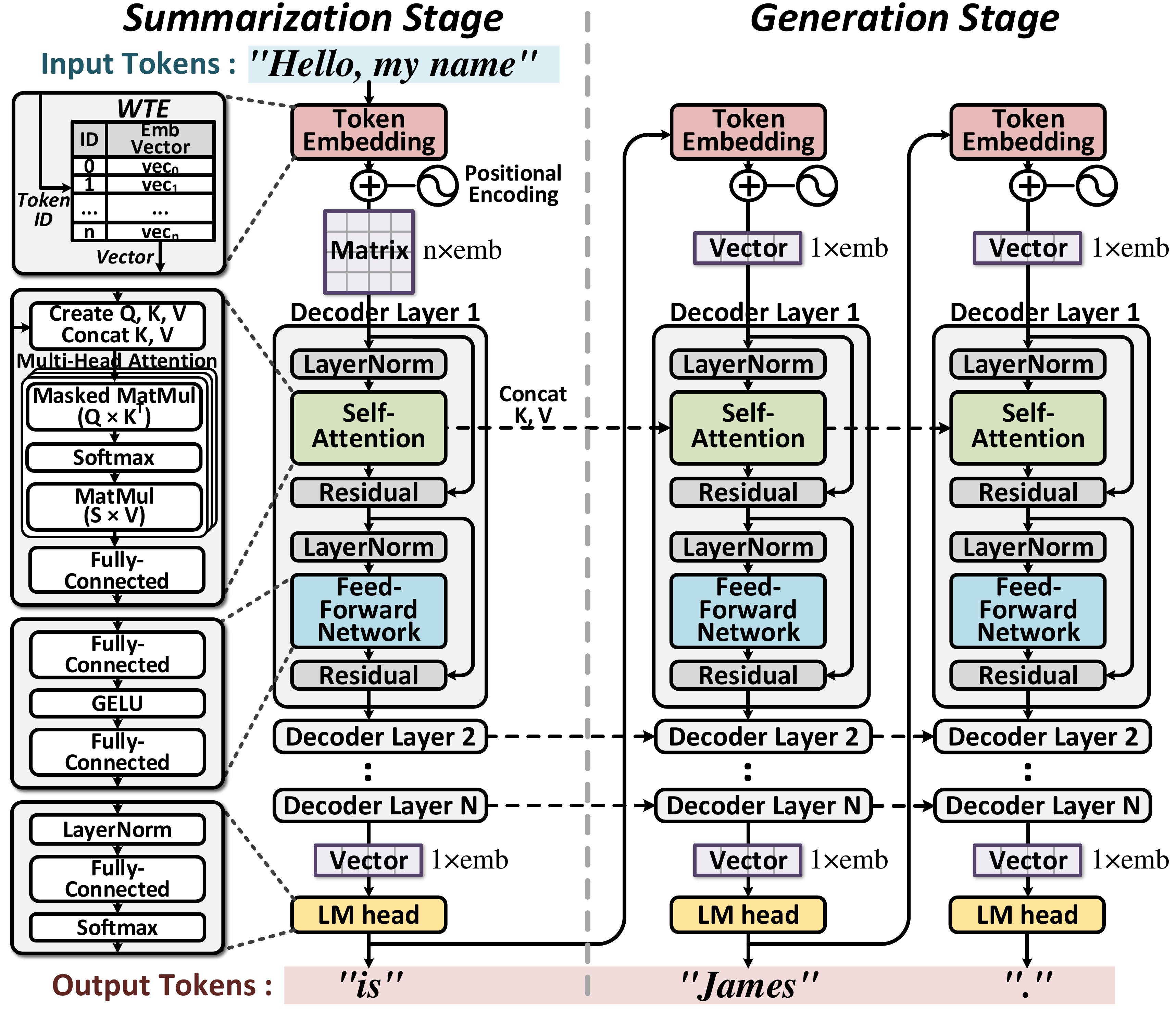} 
\vspace{-0.05in}
\caption{GPT-2 structure and illustration of summarization and generation stages in text generation.}
\label{fig-gpt_structure} 
\vspace{0.05in}
\end{figure}

% Overall Architecture of GPT
GPT is based on the transformer\cite{vaswani2017attention} structure that achieves the best accuracy in NLP. The primitive transformer has two parts, encoder and decoder, to process the input and output sequence, respectively. However, GPT includes only the decoder because it focuses on generating text (i.e., creating the next word sequence) by looking at the given context. GPT is able to remove the encoder by using an alternate method called token embedding, a process that uses pre-trained matrices in place of the encoder. Furthermore, the model size of GPT and its decoder layer is constantly increasing with more parameters and operations to gain better accuracy and sophistication in its token generation. Recently, OpenAI announced the latest GPT model, GPT-3, but the model itself is not available in the public domain~\cite{openaigpt3}. In this paper, we are based on the publicly available GPT-2 model. Note that our hardware acceleration strategies for GPT-2 are applicable to GPT-3 because it has the same model structure but with a larger size.
Figure \ref{fig-gpt_structure} shows the GPT-2's model structure and GPT-2-based text generation workload.

%%%%%%%%%%%%%%%%%%%%%%%%%%%%%%%%%%%%%%%%%%%%%%%%%%

% Overview of decoder and explain token embedding / LM head
\textbf{GPT-2 Structure}
The token embedding, located at the beginning of the decoder, is responsible for converting an input word(s) into an embedding vector. The input word is converted to the numeric token ID based on a dictionary. Then, the pre-trained matrices, word token embedding (WTE) and word position embedding (WPE), are indexed with the token ID to obtain the corresponding vectors. WTE contains token-related encoding, and WPE contains position-related encoding. The two vectors are added to get the embedding vector. LM head, located at the end of the decoder, has the opposite role to the token embedding. It converts the output embedding vector into the token ID. This process requires a matrix multiplication with the transpose of WTE and selects the token ID with the highest probability value by applying the softmax function. The selected token ID represents the generated word.

% Explain the rest of the decoder part
GPT-2 has $N$ number of decoder layers between token embedding and LM head, in which $N$ is determined by the model size. As shown in Figure \ref{fig-gpt_structure}, one decoder layer is largely divided into four operations: self-attention, feed-forward network, layer normalization, and residual. Self-attention, a type of attention for the decoder, is a key part of the transformer \cite{vaswani2017attention}; it creates the $Query$, $Key$, and $Value$ matrix to obtain the attention matrix. $Query$ is related to a current given word, while $Key$ and $Value$ represent the flow of the entire context. GPT-2 uses the multi-head attention\cite{radford2019language}, a method of dividing attention weights into $H$ columns to execute $H$ independent matrix operations in parallel. $H$ represents the number of attention heads, and this hyperparameter is increasing with the model size. Another significant operation is the feed-forward network, which is commonly used in deep neural networks. It is made up of two fully connected (FC) layers and Gaussian Error Linear Unit (GELU) activation function \cite{hendrycks2016gaussian}. Lastly, the layer normalization and residual, well-known techniques in previous works \cite{ba2016layer, he2016deep}, are placed around self-attention and feed-forward network to fine-tune the large model. 

%The first FC layer makes a vector of four times the column width for higher resolution \cite{hendrycks2016gaussian}, which is sent to the GELU function. The second FC layer restores the dimension of the vector to the original size.

% Summarization and Generation Stage
To generate tokens with the given context, GPT-2 contains a summarization and generation stage. The summarization stage takes the entire context as input, so the decoder's input dimension after the token embedding is $\texttt{n}\times\texttt{emb}$, in which \texttt{n} is the length of the context in tokens and \texttt{emb} is the length of the embedding vector; for reference, $\texttt{emb} = 1600$ for the 1.5B model. The embedding vector is fed to the decoder, which involves a series of matrix multiplications with weights of size $\texttt{emb}\times\texttt{emb}$ or larger, to produce an output matrix with the same initial dimension of $\texttt{n}\times\texttt{emb}$. Only the last row of the output matrix is processed in LM head, and the first subsequent token is generated. The $Key$ and $Value$ matrices that represent the context are also created in the summarization stage. In the generation stage, the previously generated token enters the decoder, so the input dimension is $\texttt{1}\times\texttt{emb}$. Since the tokens generated are determined by the previous context, the generation stage updates the $Key$ and $Value$ matrices by appending a row with each new input's context. For instance, if ``Hello, my name'' is the context with an input token length of 4,  $\texttt{4}\times\texttt{emb}$ $Key$ and $Value$ matrices are formed, and the first output token that represents the word, ``is,'' is generated in the summarization stage. If the requested output token length is 3, the output token enters two iterations of the generation stage, and the $Key$ and $Value$ matrices increase their row dimension by 1 after each iteration. From the decoder, the next tokens that represent ``James'' and ``.,'' are outputted sequentially. Finally, the generated tokens are detokenized altogether to form the sentence ``Hello, my name is James."

The length of the given context and the length of the output words affect the amount of computation in the summarization and generation stage, respectively, so the time spent on either stage varies for different workloads.

\textbf{GPT-2 Workload}
Completing the processes above, GPT-2 is able to generate words or sentences from the input context. 
%Therefore, GPT-2 is suitable for many workloads related to text generation. 
Some prominent text generation workloads include dialogue system and topic-to-essay generation (e.g., chatbot and article writing). Depending on the workload, the ratio of context to generation varies. For instance, the chatbot service has an average input token request of length 50, then produces an output token of length 50, having a ratio of 1:1. In contrast, the article writing application from OpenAI allows the user to input up to 50 tokens, then produces up to 150 tokens, having a wide range of ratios from 50:1 to 1:150\cite{tokenratio}. In other less widely used applications for GPT-2 like question-and-answer, the input context is much longer to generate a few-word answer. In datacenters, GPT-2 is applied for language services that tend to require more output tokens in the generation stages than the input tokens in the summarization stage.

\subsection{Parallelism in Deep Learning} \label{background_parallelism}

To train or inference large NLP models, multiple computing devices, usually GPUs, are used\cite{shazeer2018mesh, huang2019gpipe}. Among various ways to process a single model through multiple workers, the following two parallelism schemes are mainly used.

\textbf{Data Parallelism}
Data parallelism\cite{cormen1996bridging} is the method of splitting the input batch across multiple workers. The workers individually perform operations with their own batch data. This method is suitable for training but not inference because a single batch size, or non-batched input, is commonly used for inference applications that involve dynamic user requests. 
%Since more workers can continue to increase the batch size, it can be applied to models that require large training batch size.  However, machine learning models used in real industry applications are challenging to use in the inference phase due to their single batch. In addition, since one worker still processes the entire model, if the size of the model itself is very large, data parallelism would not be able to get computational advantages.

\textbf{Model Parallelism}
Model parallelism separates the model parameters across multiple workers and processes them simultaneously. It is beneficial for large models such as GPT-2\cite{radford2019language} and BERT\cite{devlin2018bert} because the size of the model allocated to each worker is reduced. 
Two widely used model parallelism schemes are pipelined parallelism \cite{tarnawski2020efficient} and intra-layer parallelism \cite{shoeybi2019megatron}. In pipelined parallelism, only one worker performs a group of operations of the model and transfer its output to different workers that process other operations of the model. The entire process is pipelined for high throughput, but the latency cannot be reduced. In intra-layer parallelism, the parallelizable operations (e.g., matrix multiplication) can be divided into multiple devices, so the execution time of operation is significantly reduced. However, synchronization may be required before certain operations that require the entire output, so the performance is dependent on the number of synchronizations and physical devices.

\section{Motivation}
\label{motivation}

%%%%%%%%%%%%%%%%%%%%%%%%%%%%%%%%%%%%%%%%%%%%%%%%%%
\begin{figure}[t] 
%\vspace{0.05in}
\centering
\footnotesize
\includegraphics[width=3.0in]{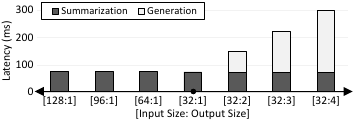}
\vspace{-0.05in}
\caption{Latency of text generation with increased number of input tokens (leftward) and output tokens (rightward) on GPU for GPT-2 1.5B model.}
\label{fig-motiv2} 
\vspace{0.1in}
\end{figure}
%%%%%%%%%%%%%%%%%%%%%%%%%%%%%%%%%%%%%%%%%%%%%%%%%%

% \vspace{-0.05in}
\subsection{Sequential Characteristic} \label{motivation_sequential}

As described in Section \ref{background_gpt}, GPT-2's summarization and generation stage have different computational property. The summarization stage needs multiple tokens simultaneously, so processing multiple tokens in parallel is advantageous. On the other hand, the generation stage is a rather sequential process that processes a single token one by one. Thus, the opportunity for parallel processing is low. In datacenters, GPU is typically used to run the text generation applications. GPU is ideal for processing large input tokens in the summarization stage with its massively parallel compute units and high memory bandwidth, but it shows significant performance degradation in the generation stage. The sequential process of the generation stage is not parallelizable, and the operations are not compute-intensive enough to effectively utilize GPU's large compute units. As a result, significant underutilization occurs in the GPU. Figure \ref{fig-motiv2} shows that each additional output token leads to a large increase in latency, 75.45ms on average, whereas each additional input token increases the latency only by 0.02ms on average for the GPU with GPT-2 1.5B model. Since text generation workloads typically create long output tokens, devising an architecture that speeds up the generation stage while maintaining high throughput in both stages is imperative.

We also examine the effect of batch size on the latency and throughput of the GPT-2. On the application level, if the datacenter chooses to batch the input of different users, the latency increases with the batch size because of the time spent gathering the input from different users \cite{fowers2018configurable}. Consequently, current datacenters prefer to run the model without fully gathering the input (i.e., non-batched input). In this case, the utilization decreases linearly by the number of unfilled inputs in the given batch, so an optimized datapath that can run a single input token with low latency is required. 

% If the datacenter chooses to run the model without fully gathering the input, the utilization decreases linearly by the number of unfilled inputs in the given batch. 

%, and the hardware may not have enough logic to handle large computations. 

\subsection{End-to-End Acceleration} \label{motivation_endtoend}

Most previous works target the acceleration of only the attention mechanism in transformers \cite{ham2020a3, ham2021elsa, zadeh2020gobo}, and few include the feed-forward network \cite{wang2021spatten}. However, GPT-2 contains additional processes: token embedding, layer normalization, residual, and LM head. Previous works disregard accelerating these additional processes because the time spent on the attention mechanism outweighs other processes. However, if these architectures are to run GPT-2 end-to-end on an existing application or service, the rest of the processes would need to be completed on the host. As the model repeats the decoder layer processing many times, extensive data transactions between the host and accelerator through the PCIe would easily become the bottleneck in the system performance. We need an accelerator that supports the entire GPT-2 process without missing any functions in the middle for practical usage.

\begin{figure}[t]
% \vspace{0.05in}
\centering
\footnotesize
\includegraphics[width=3.33in]{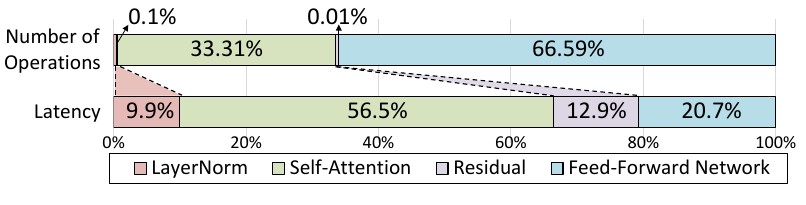}
\vspace{-0.05in}
\caption{GPT-2 latency and number of operations breakdown on GPU.}
\label{fig-breakdown} 
\vspace{0.1in}
\end{figure}

Moreover, GPT-2 contains operations that are suboptimal for the GPU to complete. Figure \ref{fig-breakdown} shows that the time spent for layer normalization and residual is at 22.8\% for the GPU even when the number of raw mathematical operations that accounts for them is extremely low at 0.11\%. This breakdown demonstrates that low-level operations of the layer normalization and residual are inefficient in the GPU. Domain-specific accelerators allow for hardware design that is specialized for these complex computations. Therefore, an alternative accelerator optimized for all GPT-2 operations is necessary.

%%%%%%%%%%%%%%%%%%%%%%%%%%%%%%%%%%%%%%%%%%%%%%%%%%
%%%%%%%%%%%%%%%%%%%%%%%%%%%%%%%%%%%%%%%%%%%%%%%%%%

%%%%%%%%%%%%%%%%%%%%%%%%%%%%%%%%%%%%%%%%%%%%%%%%%%
% Figures
%%%%%%%%%%%%%%%%%%%%%%%%%%%%%%%%%%%%%%%%%%%%%%%%%%
\begin{figure*}[t] 
% \vspace{-0.2in}
\centering
\footnotesize
\includegraphics[width=6.0in]{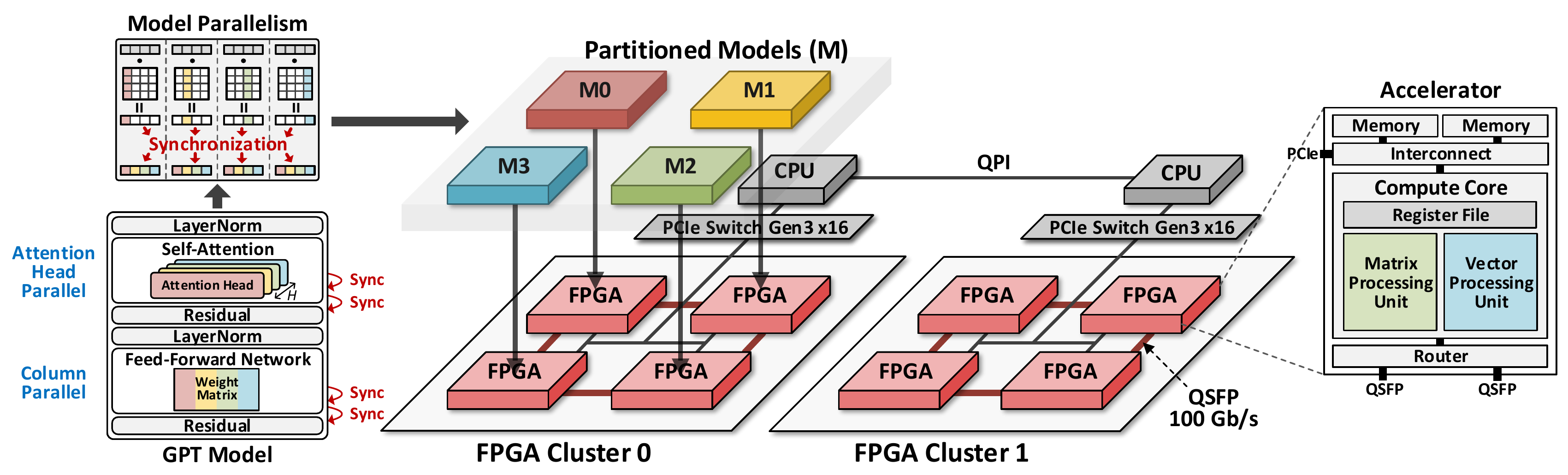}
\vspace{-0.05in}
\caption{Overall DFX appliance architecture. Left is the illustration of GPT-2 model parallelism. Center is the mapping of the partitioned models into the \sysname{} appliance. Right is the accelerator's microarchitecture.}
\label{fig-architecture} 
% \vspace{-0.15in}
\end{figure*}
%%%%%%%%%%%%%%%%%%%%%%%%%%%%%%%%%%%%%%%%%%%%%%%%%%

\subsection{Parallel Computing} \label{motivation_parallel}

GPT-2 requires massive computations with large model parameters, which suggests the need for a parallelism scheme that divides the large model into multiple nodes for parallel computation. With the growing dimensions of the GPT-2 model, a single device is not sufficient because it lacks both memory bandwidth and capacity to do the required computations. The latency is also critical in text generation workloads, so multiple devices must share the workload to reduce the overall execution time for the given input token. To address this shortcoming, a multi-device system that adopts model parallelism and efficient network is necessary to maximize the amount of parallel computation with minimal additional latency.

% \subsection{Cost} \label{motivation_cost}
% Amazon Web Service (AWS), Naver Clova, and other companies provide NLP-based services through their datacenters. Most datacenters install multitude of datacenter GPUs such as NVIDIA V100 and A100 or their server solutions such as DGX and DGX A100 to run their services. However, the total cost of ownership is extremely high for these datacenters due to the upfront cost of acquiring the hardware and the operating cost of running the hardware. Therefore, a server appliance that has low product cost and energy-efficient implementation but does not compromise on performance is highly demanded.

% \vspace{0.1in}

% \vspace{-0.1in}

\section{\sysname~Architecture}
\label{architecture}

% \vspace{-0.07in}
\subsection{Architecture Overview}
\sysname{} architecture is designed to efficiently accelerate large transformer-based language models based on FPGAs. As shown in Figure \ref{fig-architecture}, \sysname{} is a server appliance architecture that consists of dual-socket CPUs and multiple FPGAs. One CPU and a homogeneous cluster of four FPGAs form a system to compute an independent workload. Each FPGA contains one compute core for a total of four cores per cluster. The FPGAs are connected to the host CPU via PCIe Gen 3 subsystem that transfers data at 16 GB/s. The FPGA-to-FPGA communication is enabled by a  QSFP transceiver that transfers data at 100 Gb/s. As each FPGA is limited to two QSFP ports, a ring network is chosen instead of other network topologies that require more node-to-node connections. 
%We minimize the data synchronization and transfer needed between the FPGAs to only a few times per decoder layer. The internal core-to-core communication has no restrictions because it depends on the available resources. We balance the router efficiently so that the speed of the internal core-to-core communication is equal to that of the FPGA-to-FPGA communication. 
Although we chose four FPGAs per cluster, it is scalable within a server appliance with only consideration of monetary cost.

% \vspace{-0.01in}
\subsection{Homogeneous Multi-FPGA Cluster}
\label{architecture_cluster}
\begin{figure}[t] 
% \vspace{0.01in}
\centering
\footnotesize
\includegraphics[width=3.4in]{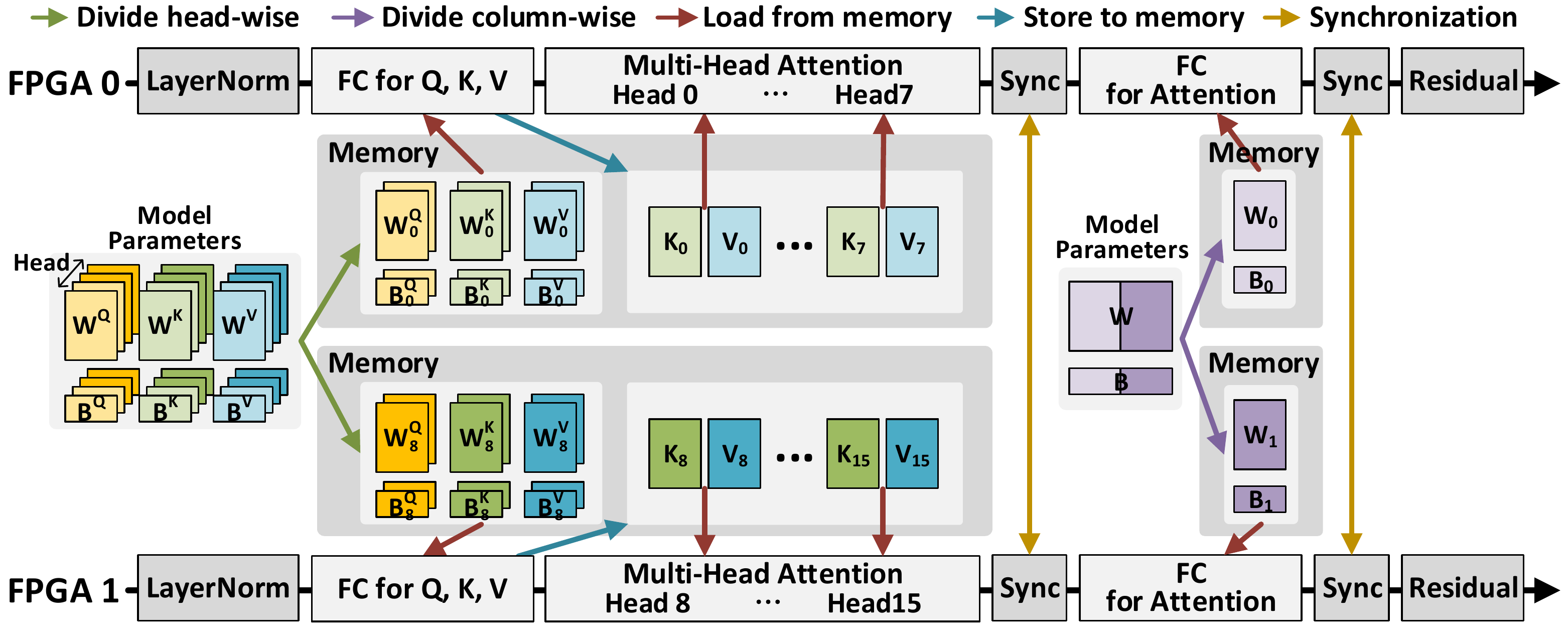}
\vspace{-0.15in}
\caption{Intra-layer model parallelism strategy on 2 FPGAs.}
\label{fig-partition} 
\vspace{0.05in}
\end{figure}

In order to efficiently process the large-scale language model, we apply model parallelism to the model parameters. In particular, we adopt intra-layer parallelism scheme\cite{shoeybi2019megatron} instead of pipelined parallelism scheme\cite{tarnawski2020efficient}. A specific intra-layer partitioning method is applied to reduce the latency of matrix operations, proportional to the number of workers, with minimal synchronization overhead. In contrast, pipelining incurs high latency because all the workers execute entirety of each operation for a single input. Moreover, if the final output result of the model is the following input (i.e., feedback loop), like in text generation, the difference in latency between the two schemes would increase linearly per decoder layer. Figure \ref{fig-partition} illustrates \sysname{}'s intra-layer parallelism strategy. The weight matrices for multi-head attention are divided head-wise, and the weight matrices for the fully-connected layer are divided column-wise into two portions (i.e., the number of FPGAs) so that each FPGA can individually work on the assigned partition. 
%Each core is assigned to operate on one of the portions. 
To this end, the partitioned matrices are stored in the memory of the FPGA where their assigned core is, and each core executes identical operations in parallel with the partitioned model parameters.
Each core then obtains the final result of the matrix operations, which is a subvector of the output vector. Each subvector is then circulated to the other FPGAs through the ring network for data synchronization. After synchronization, each core has the complete vector and is ready to proceed to the next vector operation, such as residual. Overall, we need this synchronization once during and once after self-attention and feed-forward network, which sums to a total of four times per decoder layer.
%and the partial results are gathered through the synchronization process. 
%Meanwhile, the model parameters for the rest of the operations (e.g., layer normalization and residual) are computationally inexpensive and require entirety of the data. Therefore, these model parameters are replicated in each core instead because the redistribution and post-synchronization process would require more latency than their computation. 
Hence, we minimize the data synchronization and transfer needed among the FPGAs while taking advantage of model parallelism on the predominant self-attention and feed-forward network operations. Overall, each FPGA runs identical operations on identical hardware to run GPT-2 end-to-end, so the four FPGAs form a homogeneous cluster.

%%%%%%%%%%%%%%%%%%%%%%%%%%%%%%%%%%%%%%%%%%%%%%%%%%
% Figures
%%%%%%%%%%%%%%%%%%%%%%%%%%%%%%%%%%%%%%%%%%%%%%%%%%
%%%%%%%%%%%%%%%%%%%%%%%%%%%%%%%%%%%%%%%%%%%%%%%%%%

\textbf{Memory Mapping}
The FPGA harnesses 8 GB HBM and 32 GB DDR, whose theoretical peak bandwidths are 460 GB/s and 38 GB/s, respectively. Since GPT-2 requires the partitioned model parameters frequently and in large portions, the memory bandwidth has a significant impact on the overall performance. Therefore, the weight matrices are stored in the HBM. On the other hand, input/output tokens, bias vectors, and other model parameters are stored in the DDR because these data are accessed once per few iterations of matrix operations or once per entire decoder stage (e.g., WTE and WPE), thus having a negligible effect on the overall performance. \sysname{} also utilizes the standard half-precision floating-point (FP16) model parameters to retain the inference accuracy.

% We implement two identical cores per FPGA because any other number of cores suffers from performance degradation. If only one core is implemented on a multi-die FPGA, the core becomes large, which causes placement and routing congestion problems and reduces the operating frequency. If more than two cores are implemented, each core becomes small, but the synchronization overhead increases. Therefore, we design a dual-core with custom instructions and dataflow to balance and reduce the placement and routing problems and synchronization overhead.

\subsection{Instruction Set Architecture} \label{architecture_isa}

% Features of ISA
\sysname~has a flexible and custom instruction set architecture (ISA) at the assembly language level to support the end-to-end processing for GPT-2 inference, unlike previous NLP accelerators that primarily focus on attention \cite{ham2020a3, ham2021elsa, wang2021spatten}. 
% accelerators that primarily focus on attention.

% Description of instruction type and group
\textbf{Instruction Set}
There are three types in the \sysname~ISA: \texttt{compute}, \texttt{dma}, and \texttt{router} instructions. The \texttt{compute} instructions are for running the main processing units and have the format \texttt{(type, src1, src2, dst)} with additional bits to determine if the source or destination location is off-chip memory or on-chip register file. The \texttt{dma} and \texttt{router} instructions are for controlling the DMA and the network router to move data of the given transfer size to and from the cores and have the format \texttt{(type, src, dst, xfer\_size)}.

Each instruction type is executed through the instruction chaining \cite{fowers2018configurable}, in which sequences of
dependent instructions operate with minimal stalling. Meanwhile, instructions without dependencies work in parallel; for instance, \texttt{compute} processes data, \texttt{dma} fetches data, and \texttt{router} fill the buffer with data from the peer device simultaneously for synchornization. The combination of instruction chaining and parallel execution enables continuous use of memory and communication bandwidth. Algorithm \ref{algorithm_1} shows the pseudocode of GPT-2 decoder layer using the ISA.

\textbf{Compute Instructions}
The \texttt{compute} instructions account for the majority of the core instruction set and are composed of two groups to control the main processing units: matrix instructions and vector instructions.
% using the \texttt{compute} instructions. 

Matrix instructions are for executing matrix-vector multiplication and additional functions such as GELU and reduce max. Matrix is loaded in tiles, and vectors are also loaded in portions. Any matrix-matrix multiplications are done by a sequence of matrix-vector multiplications. The description of the major matrix instructions is as follows.

%%%%%%%%%%%%%%%%%%%%%%%%%%%%%%%%%%%%%%%%%%%%%%%%%%
% GPT Decoder Block Algorithm
%%%%%%%%%%%%%%%%%%%%%%%%%%%%%%%%%%%%%%%%%%%%%%%%%%
\begin{algorithm}[t]
\footnotesize
%\scriptsize
\caption{\small{GPT-2 Decoder Layer}}
\label{algorithm_1}
\begin{algorithmic}
\item
{\bf{Input:}} $in\_emb$, input embedding vector
\\
{\bf{Output:}} $out\_emb$, output embedding vector
\\
{\bf{Parameter:}} $H$, number of attention heads 
\begin{algorithmic}[1]
\State{\textcolor{olive}{/* Layer Norm */}}
\State{$lnorm1 = LayerNorm(in\_emb, \gamma_{l1}, \beta_{l1})$}
\State{\textcolor{olive}{/* Self-Attention */}}
\State{$value = Conv1D(lnorm1, W_v, b_v)$}
\State{$key = Conv1D(lnorm1, W_k, b_k)$}
\State{$query = Conv1D(lnorm1, W_q, b_q)$}
\For{$h = 0$ \textbf{to} $H$}
\State{$mat = MaskedMM(query[h], key^T[h])$}
\State{$redu\_max = ReduMax(mat)$}
\State{$score = Softmax(mat - redu\_max)$}
\State{$attn'[h] = MM(score, value[h])$}
\EndFor
\State{$attn = Sync(attn')$}
\State{$c\_attn' = Conv1D(attn, W_a, b_a)$}
\State{$c\_attn = Sync(c\_attn')$}
\State{\textcolor{olive}{/* Residual */}}
\State{$c\_attn = c\_attn + in\_emb$}
\State{\textcolor{olive}{/* Layer Norm */}}
\State{$lnorm2 = LayerNorm(c\_attn, \gamma_{l2}, \beta_{l2})$}
\State{\textcolor{olive}{/* Feed-Forward Network */}}
\State{$ffn1' = GELU(Conv1D(lnorm2, W_{f1}, b_{f1}))$}
\State{$ffn1 = Sync(ffn1')$}
\State{$ffn2' = Conv1D(ffn1, W_{f2}, b_{f2})$}
\State{$ffn2 = Sync(ffn2')$}
\State{\textcolor{olive}{/* Residual */}}
\State{$out\_emb = ffn2 + c\_attn$}
\end{algorithmic}
\end{algorithmic}
\end{algorithm}
%%%%%%%%%%%%%%%%%%%%%%%%%%%%%%%%%%%%%%%%%%%%%%%%%%

\begin{enumerate} [leftmargin=*]
\setlength\itemsep{0.01in}
    \item \texttt{Conv1D}: This essential matrix instruction, written as the equation $Ax + b$, is used in $Query$, $Key$, $Value$ matrix generation and the feed-forward network. In this instruction, weight matrix $A$, input vector $x$, and bias vector $b$ are required for execution. \texttt{Conv1D} has a convolutional aspect in that if its input is longer than the maximum input size, the operation is performed through a sliding window.
    \item \texttt{MaskedMM}: Masked matrix multiplication (MM) has the equation, $Ax$. \texttt{MaskedMM} calculates $Query \times Key^T$, also known as the $Score$ matrix. Note that the $Query$ matrix is loaded as vectors. The masking operation puts a $-\infty$ mask on the upper diagonal elements of the $Score$ matrix to indicate that the current token is not impacted by future contexts. In combination with \texttt{Softmax}, a vector instruction, \texttt{MaskedMM} creates a lower triangular matrix and outputs the maximum value of each row.
    \item \texttt{MM} : \texttt{MM} instruction is the same as \texttt{MaskedMM} without masking. It is used in LM head for calculating logit, an intermediate value produced while converting the output embedding vector to its token ID, and in the attention layer for calculating $Score \times Value$. 
    %Multiplying $Value$ also requires the transpose unit.
\end{enumerate}

Vector instructions execute low-level vector-vector and vector-scalar operations along with \texttt{load} and \texttt{store}. They support various basic operations, including \texttt{add}, \texttt{sub}, \texttt{mul}, \texttt{accum}, \texttt{recip\_sqrt}, \texttt{recip}, and \texttt{exp}. Thus, some high-level operations such as \texttt{LayerNorm} and \texttt{Softmax} are effectively implemented by several vector instructions.

\begin{enumerate} [leftmargin=*]
\setlength\itemsep{0.01in}
    \item \texttt{LayerNorm}: Layer normalization has the equation 
    $y(x_i)=\gamma_{i} \frac{x_i-\mu}{\sigma}+\beta_{i},$ in which the $\mu$ and $\sigma$ are mean and standard deviation, and $\gamma$ and $\beta$ are weight and bias vectors, respectively.
    Calculating the mean requires \texttt{accum} and \texttt{mul} instructions, and the standard deviation  requires \texttt{recip\_sqrt} in addition. The formula is then executed by the \texttt{sub}, \texttt{mul}, and \texttt{add} instructions. The parameters are fetched to the register file through the \texttt{load} instruction.
    \item \texttt{Softmax}: Softmax has the equation
    $y(x_i)=\frac{e^{x_i}}{\sum_j{e^{x_j}}},$ 
    in which $j$ is the number of elements in the row. This operation can be performed with basic vector instructions, such as \texttt{exp}, \texttt{add}, and \texttt{accum}. The summation is similar to calculating the mean in the \texttt{LayerNorm}. The division is substituted by the \texttt{recip} and \texttt{mul} instructions. %\vspace{-0.1in}
\end{enumerate}

%%%%%%%%%%%%%%%%%%%%%%%%%%%%%%%%%%%%%%%%%%%%%%%%%%
% Figures
%%%%%%%%%%%%%%%%%%%%%%%%%%%%%%%%%%%%%%%%%%%%%%%%%%
\begin{figure}[t] 
% \vspace{0.01in}
\centering
\footnotesize
\includegraphics[width=3.33in]{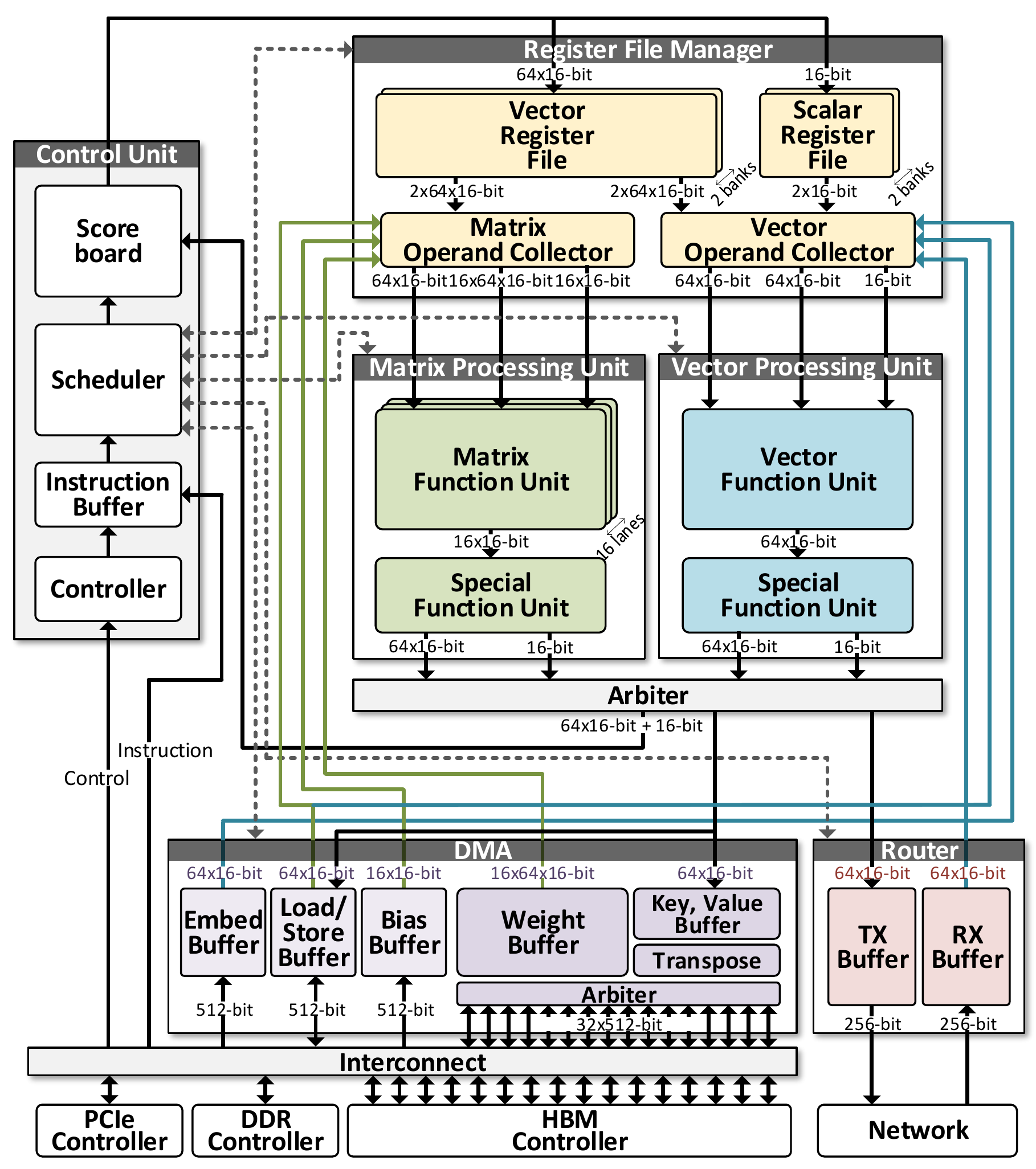}
\vspace{-0.05in}
\caption{DFX compute core microarchitecture.}
\label{fig-core} 
\vspace{0.05in}
\end{figure}
%%%%%%%%%%%%%%%%%%%%%%%%%%%%%%%%%%%%%%%%%%%%%%%%%%

%\subsection{Quantization} \label{architecture_quantization}

%To resolve the memory-intensive problem of GPT-2 inference, we quantize the model parameters from half-precision floating-point (FP16) to 8-bit minifloat (FP8) and preloaded into the HBM. This quantization method enables 2$\times$ the data transfer to and from the HBM and thus 2$\times$ the number of computations per cycle. FP8 is composed of 1-bit sign, 4-bit exponent, and 3-bit mantissa (1-4-3). Within the core, however, \sysname{} dequantizes FP8 back to standard FP16 (1-5-10) because having FP16 processing units and on-chip memory prevent further decrease in accuracy that occur from truncation or rounding. We tested different configurations of FP8 (e.g., 1-5-2, 1-4-3, and 1-3-4) and determine that 1-4-3 results in marginally better accuracy on average across the benchmarks, but our hardware is flexible to dequantize any data type configurations depending on the application. We do not consider fixed-point with higher precision than FP8 because NLP models have processes like layer normalization that require a large dynamic range \cite{tambe2021edgebert}. We also do not consider integer quantization with a higher quantization ratio because of the significant accuracy loss that occurs with such drastic compression. 

% \vspace{0.1in}

% \vspace{-0.01in}

\section{Microarchitecture}
\label{microarchitecture}

Figure \ref{fig-core} shows the proposed compute core's microarchitecture, which mainly consists of matrix processing unit and vector processing unit. The primary goal of microarchitecture is to efficiently process text generation workloads that have sequential processes with non-batched input. In the following subsections, we explain the details of microarchitecture.

%Figure \ref{fig-core} shows the proposed compute core's microarchitecture, which consists of a control unit, DMA, register file manager, processing units, and router. In the following subsections, we explain the details of each unit.

% \subsection{Dataflow} \label{architecture_dataflow}
% The dataflow of \sysname{}, shown in Figure \ref{fig-core}, is designed based on the characteristics of the GPT model to efficiently support the ISA. Memory and network interfaces are responsible for sending data to and receiving data from the core via the AXI-Stream protocol. The memory interface sends data from the HBM and DDR to the core through the direct memory access (DMA) that supports a custom tiling scheme, and the network interface sends synchronization data between cores and between devices through the lightweight router.

% \vspace{-0.02in}
\subsection{Control Unit} \label{microarchitecture_control}
The control unit contains logic to control the overall flow of data by keeping track of the state of each unit and arbitrating which modules to run. It is composed of the controller, scheduler, and scoreboard.

\textbf{Controller}
The controller's main job is to receive the start signal and system configuration from the host. %via the AXI-Lite interface, decode them, and pass them to the scheduler. 
The system configuration includes the core ID and the number of cores in the system, and the number of decoder layers and tokens that the system needs to run on. These parameters determine the behavior of each core. 
The core ID and the number of cores direct the corresponding core on which section of the model weights to work on and which peer device to receive from and transmit to. The number of decoder layers determines when single token processing completes, and the number of input and output tokens determines when the entire service completes. Since a different portion of the HBM needs to be accessed for each layer, the layer number designates the address the DMA needs to access. 
The token number is used specifically for knowing where to mask during \texttt{MaskedMM}.
Lastly, the controller returns the done signal back to the host once the entire GPT-2 operation finishes.
 
\textbf{Scheduler}
The scheduler receives the decoded system configuration from the controller and instructions from the instruction buffer. The scheduler contains multiple finite state machines for each instruction type that checks the status of the DMA, processing units, register file, and the router to decide whether to run or wait on each instruction type. The chosen instruction is sent to the scoreboard for the last dependency check with the running instruction.

\textbf{Scoreboard}
The register file needs to check for dependencies to run instructions based on the chaining method. Since a sequence of instructions may cause data hazards, the scoreboard monitors source and destination addresses. The scoreboard uses a RAM to represent the address space and marks the current instruction's address with a \texttt{stale} bit when in execution and with a \texttt{valid} bit when in writeback. If the source and destination addresses overlap, %or other hazards occur, 
the next instruction stalls until the current computation finishes. %The instruction that passes the scoreboard executes.

\subsection{Direct Memory Access} \label{microarchitecture_dma}
The DMA, which contains the read and write interface, serves a vital role in distributing the data that is transferred at high bandwidth. To maximize the bandwidth of the HBM, the DMA's read and write interface is connected to all 32 HBM channels and handles single-channel data bitwidth of 512 bits at 200 MHz for a total of $32 \times 512$ bits per cycle. The DMA stores and loads tiled weights, $Key$, and $Value$ to and from the HBM optimized for the matrix operation. In our dataflow, the output $Value$ needs to be transposed when being written, so a transpose unit is placed in the DMA. Besides the HBM channels, the single DDR channel is also accessible by the DMA. 
The input token, bias, WTE, and WPE are read from DDR to their corresponding buffers in the DMA. The final output token is also written from the DMA to DDR. 
% Because the weights and biases cannot be reused in matrix-vector multiplication due to no input batching, they are buffered in the DMA and streamed into the processing units for the computation with the preloaded input.
Moreover, weights and biases cannot be reused in matrix multiplications due to no input batching, so they are buffered in the DMA and streamed into the processing units for the computation with the preloaded input.

% %%%%%%%%%%%%%%%%%%%%%%%%%%%%%%%%%%%%%%%%%
% \begin{figure}[t]
% \centering
% \includegraphics[width=2.5in]{Figures/KeyValue.pdf}
% \vspace{-0.15in}
% \caption{Memory requirement for key and value on various GPT-2 models.}
% \label{fig-keyvalue}
% \vspace{0.05in}
% \end{figure}
% %%%%%%%%%%%%%%%%%%%%%%%%%%%%%%%%%%%%%%%%%

\textbf{Tiling Scheme} \sysname{} uses an optimized tiling scheme that maximizes the number of computations and throughput in the memory-intensive generation stage while retaining performance in the summarization stage. To process a single token in the generation stage, a large amount of weights needs to be read from the HBM for the matrix multiplication. Therefore, the weights are tiled in the HBM, and the DMA reads the tiled weights at the maximum read bandwidth of $32 \times 512$ bits per cycle. This dimension can be rearranged to $d \times l \times BW_{data}$ weight bits, in which $d$ is the tile dimension, $l$ is the number of lanes, and $BW_{data}$ is the data bitwidth. The number of lanes is the number of columns in a tile that can be computed in parallel by the matrix function units (see Section \ref{microarchitecture_pu}).
Since \sysname{} uses FP16 data as its data type, $BW_{data}$ is set to 16. We propose a model-and-hardware-aware tiling scheme that finds 1) the optimal $d$ and $l$ value for loading the weights of size $\texttt{emb}\times\texttt{emb}$ or larger to the DMA and 2) the effective loading direction that translates to the order of matrix-vector multiplication.

\begin{figure}[t] 
% \vspace{0.01in}
\centering
\footnotesize
\includegraphics[width=3.33in]{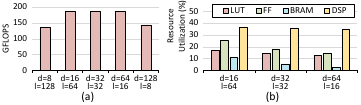}
\vspace{-0.03in}
\caption{Design choices in tile dimension and lane number and their impact on (a) multi-head attention performance and (b) resource utilization for the matrix processing unit.}
\label{fig-dse} 
\vspace{0.02in}
\end{figure}

We conduct a design space exploration to determine that the optimal $d$ and $l$ is 64 and 16, respectively. We evaluate the performance with different $d$ values, and the corresponding $l$ values are chosen to maximize the memory bandwidth with FP16 data: $(d,l)=\{(8,128), (16,64), (32,32), (64,16), (128,8)\}$. Figure \ref{fig-dse}(a) shows that $(d,l)=\{(16,64),(32,32), (64,16)\}$ have the best performance on multi-head attention with negligible difference. Since attention head dimension, $H$, for the state-of-the-art models, GPT-2 and GPT-3, is around 64 \cite{radford2019language, brown2020language}, we examine that $d>64$ leads to underutilized compute units and thus lower performance when computing the multi-head attention. Specifically, performance degradation occurs when computing $Query \times Key^T$ because $Key^T$ has $H$ rows which is smaller than $d$ when $d>64$. Similarly, $l>64$ also leads to performance degradation when calculating $Score \times Value$ because $Value$ has $H$ columns. Then, we synthesize the hardware resource utilization required for the three choices to find that $d=64$ requires the least amount of hardware resource as shown in Figure \ref{fig-dse}(b). For $d=16$ and $d=32$, larger $l$ is required to maintain the same number of operations. With larger $l$, the number of MAC remains unchanged, but the resources in the matrix processing unit (e.g., accumulator, operators in the special function unit, and the control logic) increase linearly. Therefore, we standardize the hardware with $d=64$ and $l=16$.

%The reason for not using $d=32$ or $d<64$ is that the vector register file is designed to match the dimension of the matrix function unit (MFU) and vector function unit (VFU), so $d=32$ would cause the latency of VFU to double. Note that the latency of MFU in the case of $d=32$ would not be affected because the number of lanes can be proportionally increased to $l=32$. After choosing $d=64$, we choose $l=16$, which maximizes the memory bandwidth usage.

%while staying within the resource constraint. Since $l$ determines the output vector dimension and also the number of partial sums, larger $l$ equates to greater number of intermediate buffers required, which can exceed the number of available hardware resources.

%%%%%%%%%%%%%%%%%%%%%%%%%%%%%%%%%%%%%%%%%%%%%%%%%%
% Figures
%%%%%%%%%%%%%%%%%%%%%%%%%%%%%%%%%%%%%%%%%%%%%%%%%%
%%%%%%%%%%%%%%%%%%%%%%%%%%%%%%%%%%%%%%%%%%%%%%%%%%

%%%%%%%%%%%%%%%%%%%%%%%%%%%%%%%%%%%%%%%%%%%%%%%%%%
\begin{figure}[t] 
\centering
\footnotesize
\includegraphics[width=3.15in]{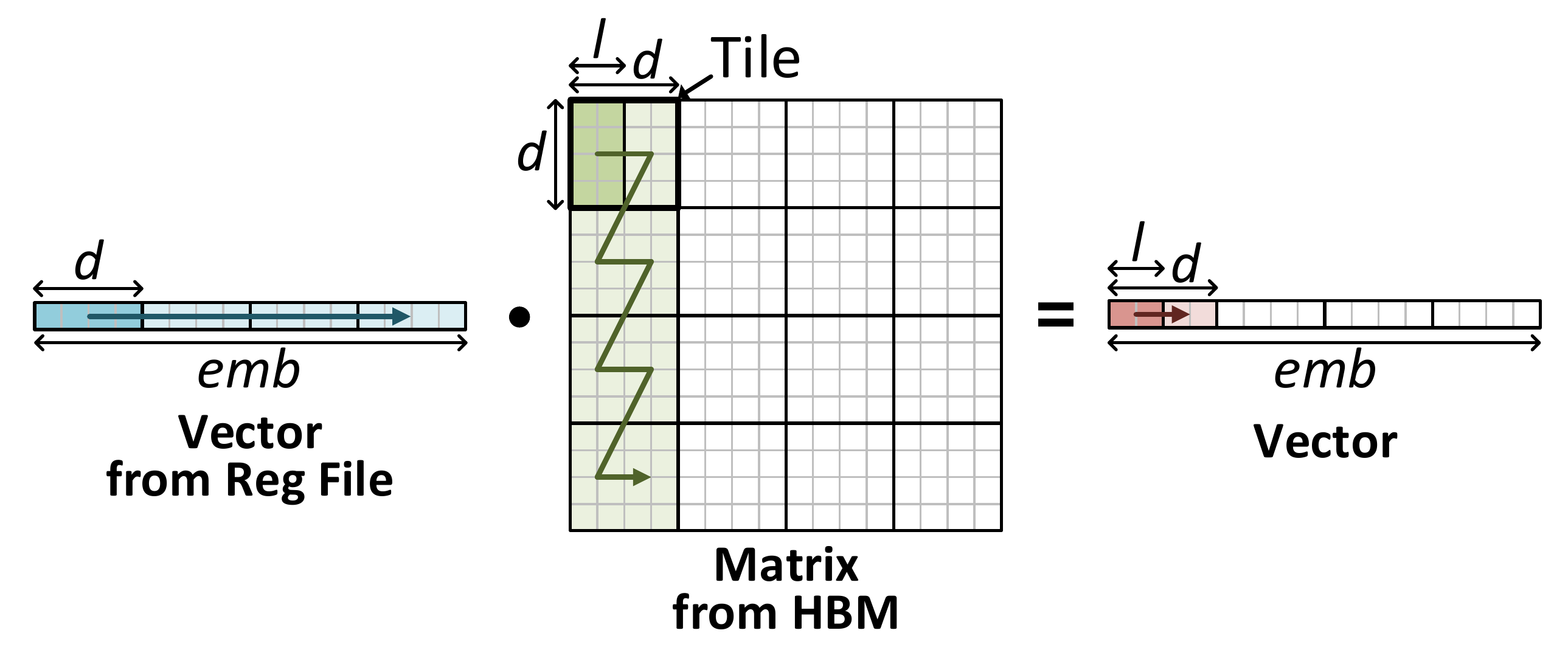}
\vspace{-0.05in}
\caption{Illustration of tiling scheme for matrix-vector multiplication.}
\label{fig-tiling} 
\vspace{0.1in}
\end{figure}
%%%%%%%%%%%%%%%%%%%%%%%%%%%%%%%%%%%%%%%%%%%%%%%%%%

Furthermore, the DMA loads $d\times l$ weights in the horizontal direction to fill in a tile, and moves to the tile below, as shown in Figure \ref{fig-tiling}. Moving in the horizontal direction maximizes input reuse, but it requires a significant number of buffers to store the partial sums that are produced when the input of length $d$ iterates across many columns of the weight matrix. As the core's requirements of deep pipelining and other buffers cause a shortage of on-chip memory, completing the horizontal direction is infeasible. The vertical direction decreases the number of buffers to one, but it removes input reuse. The inability to reuse the input increases the amount of register file access, which decreases the throughput. Therefore, the zigzag direction with a tile size of $d \times d$ balances hardware resource and data reuse for maximum performance. 
%Figure \ref{fig-tiling} shows the proposed tiling scheme.
%The controller skips the computation for tiles formed entirely of $-\infty$ masks or zero attention score because they have no impact on the output attention.

% \textbf{Transpose Scheme} In attention operation, there are key and value matrix. And their data size is too large, so it is infeasible to store in on-chip memory as shown Figure \ref{fig-keyvalue}. therefore, off-chip memory is necessary to store key and value.
% In addition, they need matrix transpose. conventional transpose scheme is loading data into on-chip memory and perform transpose as shown in Figure \ref{fig-transpose}. But, this approach occurs latency in text generation. So we put transpose unit in between register file and memory controller to remove latency.

\textbf{Transpose Scheme} 
In a standard attention operation, $Key$ needs to be transposed, but in our dataflow, the multiplicand $Value$ needs to be transposed because the read from the HBM is column-wise and the write is row-wise; therefore, the intermediate matrices are transposed by default when loaded to the DMA.
$Value$ requires high memory capacity (e.g., 0.31 MB per token for 1.5B model), so the conventional transpose scheme of transposing the entire matrix in the on-chip memory is inefficient. %The latency is also high because the transpose needs to be done in a stream after loading the entire matrix through sequential DMA read. 
To address this issue, \sysname{} transposes the $Value$ matrix while its partial tiles are being written to the off-chip memory, instead of doing it when they are read. %which reduces the memory requirement. 
%The latency problem remains, but the transpose latency can be completely hidden in this scheme as long as the transpose is finished before the transposed matrix is needed. 
%The issue of transpose latency exists, but it can be completely hidden with the following scheme.
The long latency of transpose can be completely hidden by changing the computation order.
Based on GPT's order of operation, $Value^{T}$ is needed after $Query$, $Key$, and $Value$ are generated. Therefore, we rearrange the \sysname{} instructions so that $Value$ is calculated earlier than $Query$ and $Key$. This rearrangement guarantees a sufficient period of time for the $Value$ transpose, while $Query$ and $Key$ are being generated. %Since transpose needs to happen in a smaller dimension than the channel bitwidth, AXI-write strobe is used to write bytes of data to the channel. 

% Since latency is critical in text generation, \sysname{} transposes the $Value$ data when it is being written to the off-chip memory, rather than reading it later.
% %\sysname{} transposes the data earlier before DMA write. 
% Since a vector is sequentially stored into memory and used only after the entire matrix is written, the latency is hidden, and the memory requirements are low. 

%It also has a double buffer and generates the vector-related microcodes in runtime to get the same advantages as the matrix operand collector.

%%%%%%%%%%%%%%%%%%%%%%%%%%%%%%%%%%%%%%%%%
\begin{figure}[t]
% \vspace{0.01in}
\centering
\includegraphics[width=3.33in]{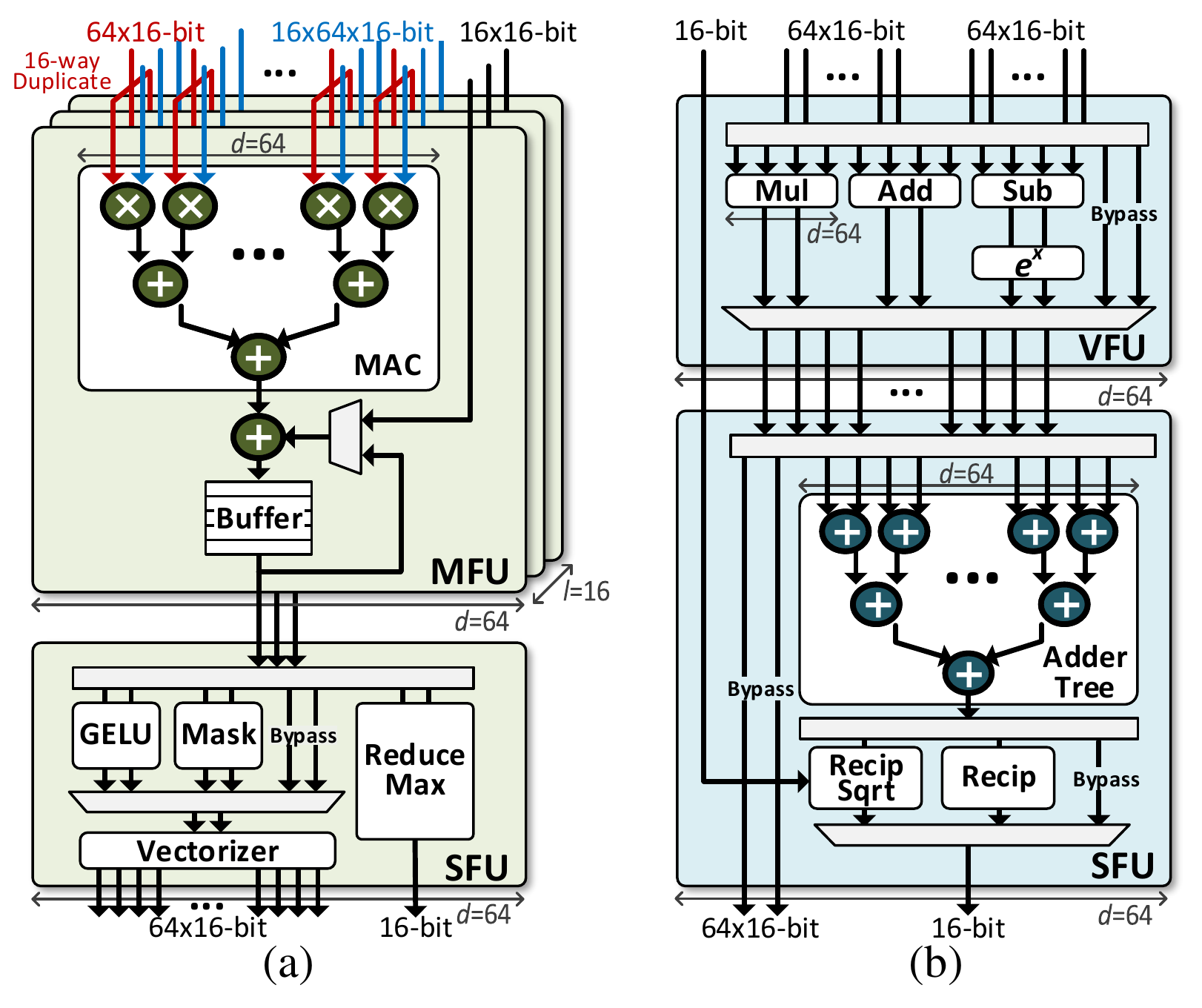}
\vspace{-0.1in}
\caption{\sysname{} processing units. (a) Matrix processing unit. (b) Vector processing unit.}
\label{fig-pu}
% \vspace{0.05in}
\end{figure}
%%%%%%%%%%%%%%%%%%%%%%%%%%%%%%%%%%%%%%%%%

\subsection{Processing Units} \label{microarchitecture_pu}
The \sysname{} core has two processing units, matrix processing unit (MPU) and vector processing unit (VPU), as shown in Figure \ref{fig-pu}. These processing units are designed to execute main mathematical operations required for the end-to-end acceleration of GPT-2, and they fully exploit parallel computing and hardware resources. The two processing units consist of four main functional units, matrix function unit and vector function unit, each accompanied by a special function unit, which all consist of FP16 operators. The functional units are composed of deep and diversified pipelines for maximum throughput and utilize bypasses at each sub-computation to asynchronously execute instructions at low latency.

\textbf{Matrix Function Unit} The matrix instructions operate on the matrix function unit (MFU). Its primary workload is matrix-vector multiplication. The MFU contains a tree-based multiplier-accumulators (MACs) that take vectors of $d$ dimensions as input. The unit is also composed of $l$ lanes, which means $l$ number of tree-based MAC hardware are in parallel. Input remains constant throughout the lanes, but $l$ different multiplicands from different columns of the weight matrix are passed to each lane, so $d \times l$ multiplications are done in parallel. %which aligns with the custom tiling scheme. 
The products in each lane are then passed to the parallel adder tree of depth $log_2(d)$ to calculate the partial sum. 
% A new partial sum is calculated every cycle because of the deep pipelining. Overall, the partial sums from the lanes are stored in a buffer, loaded and added with the new partial sums, and repeated until the final results are calculated. 

Each of the FP16 multiplier and adder is mapped to one digital signal processing slice (DSP) and two DSPs, respectively. The multiplier takes 6 cycles, and the adder takes 11 cycles. The MFU uses a total of $3\times(d \times l)$ DSPs: $d \times l$ DSPs for the multipliers, $2 \times (d-1) \times l$ DSPs for the adder trees, and $2 \times l$ DSPs for scalar additions.
In our case, $d$ and $l$ are set to 64 and 16 as explained in Section~\ref{microarchitecture_dma}, comprising of 3072 DSPs for the MFU.

%A total of $2\times d \times l-1$ DSPs are used for the tree-based MACs: $d \times l$ DSPs for the multipliers and $d \times l-1$ DSPs for the adder trees. The MFU uses $l$ more DSPs for adding scalars.

%The tile is optimized to use the maximum number of MAC hardware thus the maximum number of DSPs. 
% Because the GPT-2 model is memory-bound, the increase in the number of DSPs used yields a linear increase in performance.

\textbf{Vector Function Unit}
The vector instructions operate on the vector function unit (VFU). VFU is a floating-point arithmetic logic unit (ALU) that supports element-wise vector operations. Specifically, the VFU supports addition, subtraction, and multiplication of two vectors of $d$ dimension. 
Similar to the MFU, DSP is used for all VFU operations. Addition, subtraction, multiplication, and exponential operation take 11 cycles, 11 cycles, 6 cycles, and 4 cycles, respectively. Exponential operation uses two DSPs, and other operations use one DSP each. No instruction requires more than one ALU operation, so all instructions are completed in the shortest possible cycles without synchronization. Additionally, VFU supports bypass to reduce unnecessary computational cycles. For instance, load and store instructions do not require any computation, so the data can skip the execution stage. As VFU has a bypass path that directly connects the input and output ports, the load and store instructions take only one cycle. The data hazards that occur with such asynchronous dataflow are handled in the scoreboard.

%More complex bypass logic is located in the special function unit, which will be explained in the corresponding section. 

% Separate yet redundant multiplier is used in both MFU and VFU for two reasons: 1) the multiplication in the VFU is only for vector-scalar multiplication or vector-self multiplication, thus adding control for such exceptions in the MFU would become an overhead, and 2) the special function (e.g., reduce mean and reduce sum) that is subsequent to VFU’s multiplier overlaps with that of the other ALU operations in the VFU, so larger redundancy would have occurred in the special function unit if the multiplier only existed in the MFU.

\textbf{Special Function Units}
The special function units (SFUs) handle the nonlinear functions in GPT-2. The output from the MFU and VFU is passed into SFU\textsubscript{M} and SFU\textsubscript{V}, respectively. The SFUs use the combination of DSP, combinational logic, and the lookup table method for optimal hardware utilization. 

SFU\textsubscript{M} is responsible for executing computations that follow matrix-vector multiplication, such as 
%and are required for the matrix instructions. It is composed of 
masking, GELU, vectorization, and reduce max. The masking unit creates a lower triangular matrix based on the tile information, in which elements above the diagonal of the output matrix are masked with the closest representable value to $-\infty$, which is eventually zeroed out after softmax. For the division needed for dividing the result with the number of attention heads, a scalar constant, we use a multiplier instead to save hardware resources. To support GELU activation function with the equation $y(x)=0.5x(1+tanh[\sqrt{2/\pi}(x+0.044715x^3)]),$ the lookup table is used with linear approximation. We sample 2048 inputs that achieve a mean squared error of 0 in half-precision floating-point and choose $[-8, 8]$ as the range because the slope converges on either side at this range. Linear approximation is sufficient for GELU that has piecewise linear characteristics, and it reduces the hardware overhead of supporting complex mathematical operations. The vectorizer uses an asymmetric buffer to concatenate outputs to match the tiling, and it is placed after the above modules to increase hardware reuse. Lastly, the reduce max unit, which finds either max or argmax value of the given vector, is designed using a parallel tree of comparators.

% Linear approximation samples $n$ inputs of equal interval from a set range,  $[min, max]$, and calculates the output of the GELU function at those input value. Using the inputs and outputs of two consecutive points, the slope and y-intercept of each interval are found, which are stored in the lookup table. At runtime, the address that represents the closest value to the given input is accessed, and the linear equation is applied to get the approximated GELU function value. We choose $[-8, 8]$ because the slope converges on either side at this range, and we sample 2048 inputs that achieves mean squared error of 0 in half-precision floating-point. Linear approximation is sufficient for GELU that has piecewise linear characteristics, and it removes the hardware overhead of supporting the complex mathematical operations in GELU.

SFU\textsubscript{V} is responsible for computations that follow the vector operations by VFU;
%which are required in the vector instructions. It is composed of an adder tree, accumulator, reciprocal, multiplier, scalar adder, and reciprocal square root module. 
they are accumulation followed by scalar reciprocal, multiplication, addition, and reciprocal square root. An adder tree is in the SFU instead of in the VFU because VFU supports instructions that only require vector outputs. The rest of the functions are provided by the floating-point DSP. Similar to SFU\textsubscript{M}, SFU\textsubscript{V} uses a multiplier to divide a constant value of embedding size.

Both SFUs also utilize bypasses so that operations that do not require certain hardware in the dataflow can skip the hardware without cycle penalties. %For instance, Figure \ref{fig-pu} shows how the result of the adder tree in SFU\textsubscript{V} can be sent directly to the multiplexer without going through the pipeline stages that match the latency of the nearby reciprocal and reciprocal square root modules. 
% This asynchronous dataflow is more likely to cause data hazards, so the scoreboard is reused with negligible control overhead for bypass. 
The ordering of GPT-2 operations can be conveniently alternated with the use of matrix and vector instructions, which is advantageous in speeding up the sequential generation.

% However, the scoreboard is already required to handle other hazards in GPT-2's innate instruction sequence, so negligible control overhead is added to support bypass. 

\subsection{Register File Manager} \label{microarchitecture_registerfile}
The register file manager contains on-chip memory structures or register files for storing a multitude of FP16 data prior to and post computation in the processing units. We have two types of register files: vector and scalar register files. These register files are responsible for communicating with the memory interface via the DMA and network via the router.
%to read and write data from the host and peer devices in an organized manner. 
The register file manager also contains operand collectors that generate and collect the processing unit instructions and determine which register file data to access based on the instructions. 

%\textbf{Vector and Scalar Register File}
%The vector and scalar register files store only portions of intermediate results that maximize reuse because on-chip memory capacity is limited. Vector and scalar register files have bitwidth of $2 \times 64 \times 16 $-bit and $2 \times 16 $-bit, respectively. They both have two banks and a depth of 2048. In each register file, two banks are activated simultaneously to read a set of two data to the processing units per cycle. Sending two sets of data is essential for maximum throughput because two sources are required per instruction. The depth is chosen based on the maximum number of data the register file needs to hold at once during the GPT-2 process.

%, which happens in LM head after processing matrix multiplication with the transpose of WTE. 

\textbf{Matrix Operand Collector}
The matrix operand collector generates matrix microcodes based on the instruction mentioned in Section \ref{architecture_isa} during runtime. %and holds these signals until the ready signal is given by the control unit's scoreboard. 
The runtime generation of microcodes decreases the amount of instruction transfer from the host. The matrix operand collector passes these microcodes and operands, such as the input vector, weight matrix, and bias vector, to the MPU for execution. It reads a single input vector from the vector register file while taking the weight and bias from the DMA buffer. It counts the tiling order and allocates the corresponding input and weight to the MPU. The identical input vector is broadcasted to the parallel hardware, while different weights and biases are distributed to each vector lane. In addition, the double buffer is used for all operands to reduce the latency and obtain high throughput.

\textbf{Vector Operand Collector}
The vector operand collector generates the microcodes the VPU to execute vector instructions, similar to the matrix operand collector. Since the VPU needs various operand types, the vector operand collector can read both vector and scalar register files. It can also access the DMA and network router buffers to perform the DMA or network instructions (i.e., load, store, and synchronization). 

%%%%%%%%%%%%%%%%%%%%%%%%%%%%%%%%%%%%%%%%%%%%%%%%%%
% Figures
%%%%%%%%%%%%%%%%%%%%%%%%%%%%%%%%%%%%%%%%%%%%%%%%%%
\begin{figure}[t] 
% \vspace{0.01in}
\centering
\footnotesize
\includegraphics[width=3.33in]{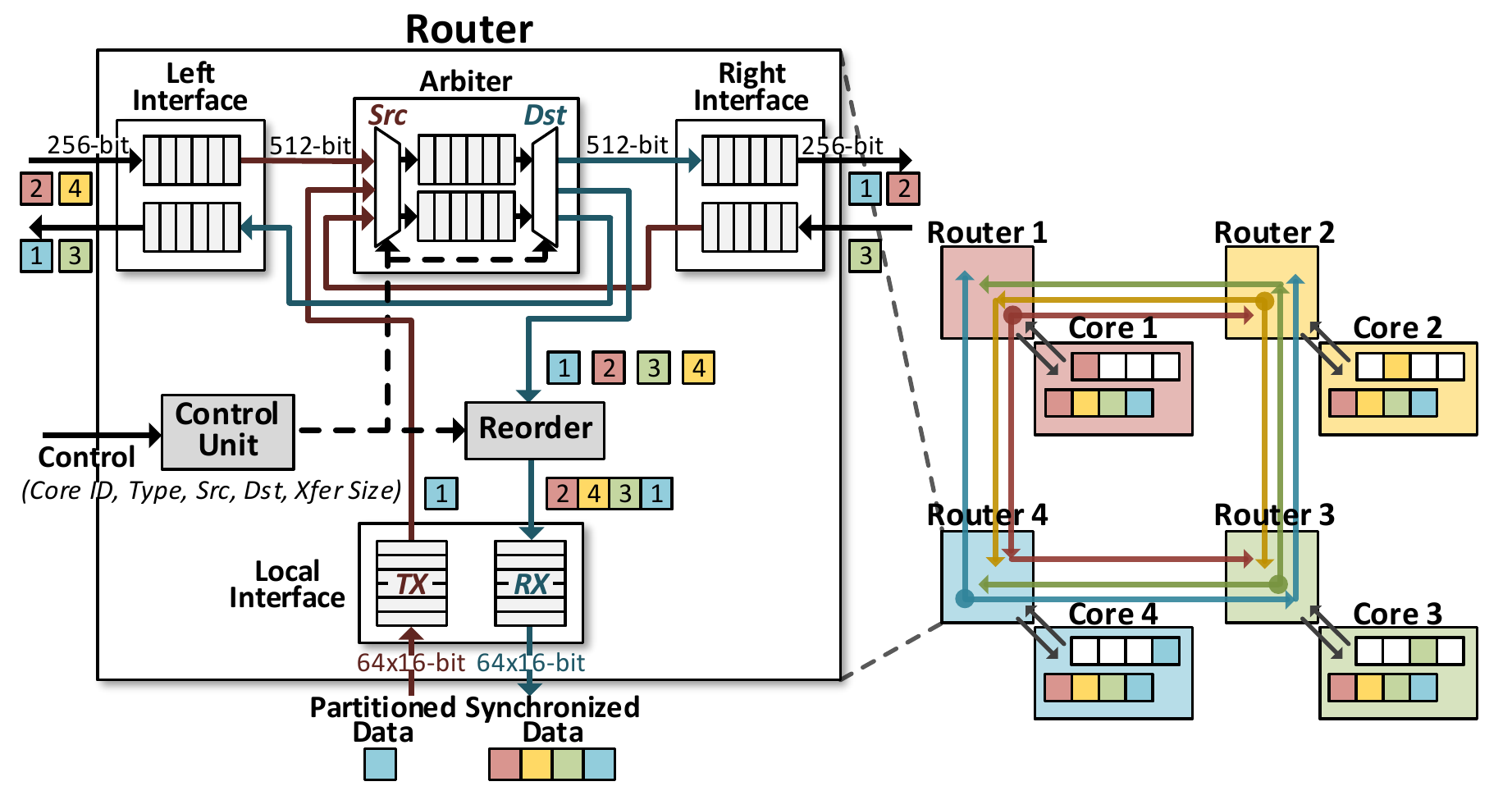} 
\vspace{-0.05in}
\caption{Illustration of data synchronization with lightweight router.}
\label{fig-router} 
\vspace{0.05in}
\end{figure}

%Lightweight router for communication of core-to-core and peer-to-peer.
%%%%%%%%%%%%%%%%%%%%%%%%%%%%%%%%%%%%%%%%%%%%%%%%%%

\subsection{Router} \label{microarchitecture_router}
The multi-FPGA network is enabled by the lightweight router. Each core utilizes the router to synchronize the data in the register files with every other core in peer devices across the ring network. Figure \ref{fig-router} shows the router structure and the data synchronization. The router seamlessly transfers $64 \times 16 $ bit data to fetch the output vector from the processing units and pass it peer-to-peer (inter-device). The router contains a control unit that indicates which device's core to communicate with, buffers to hold the transmitted and received vectors, and a reorder module that uses the core ID to organize the data order to be identical in every core. Unlike general routers, this router does not contain additional logic for packet encoding or decoding. The synchronization is necessary after executing a \texttt{Conv1D} instruction in the self-attention and feed-forward network because model parallelism leads to each core only computing a portion of the output matrix's row, and the next operation like layer normalization and residual requires the entire row. 

The network's peer-to-peer communication is enabled by Aurora 64b/66b IP \cite{aurora}. The Aurora IP implements a light link-layer protocol for high-speed serial communication between two devices. The protocol uses 64b/66b encoding, which requires a low resource cost with only 3\% transmission overhead. Consequently, the router provides a lightweight communication interface between supported devices, resulting in low latency data communication.

% The network's peer-to-peer communication is enabled by using two methods: kernel to kernel streaming (K2K) and via Aurora 64b/66b IP \cite{aurora}. The K2K performs streaming between two cores using the AXI-Stream interface, and the communication overhead is negligible. The Aurora IP implements a lightweight link-layer protocol for high-speed serial communication between two devices. The protocol uses 64b/66b encoding, which requires low resource cost with only 3\% transmission overhead. Consequently, the router provides lightweight communication interface between supported devices, resulting in low latency data communication.

% \vspace{0.1in}

% \vspace{-0.05in}

\section{Appliance Implementation} \label{implementation}

\begin{figure}[t]
% \vspace{-0.01in}
\centering
\includegraphics[width=3.33in]{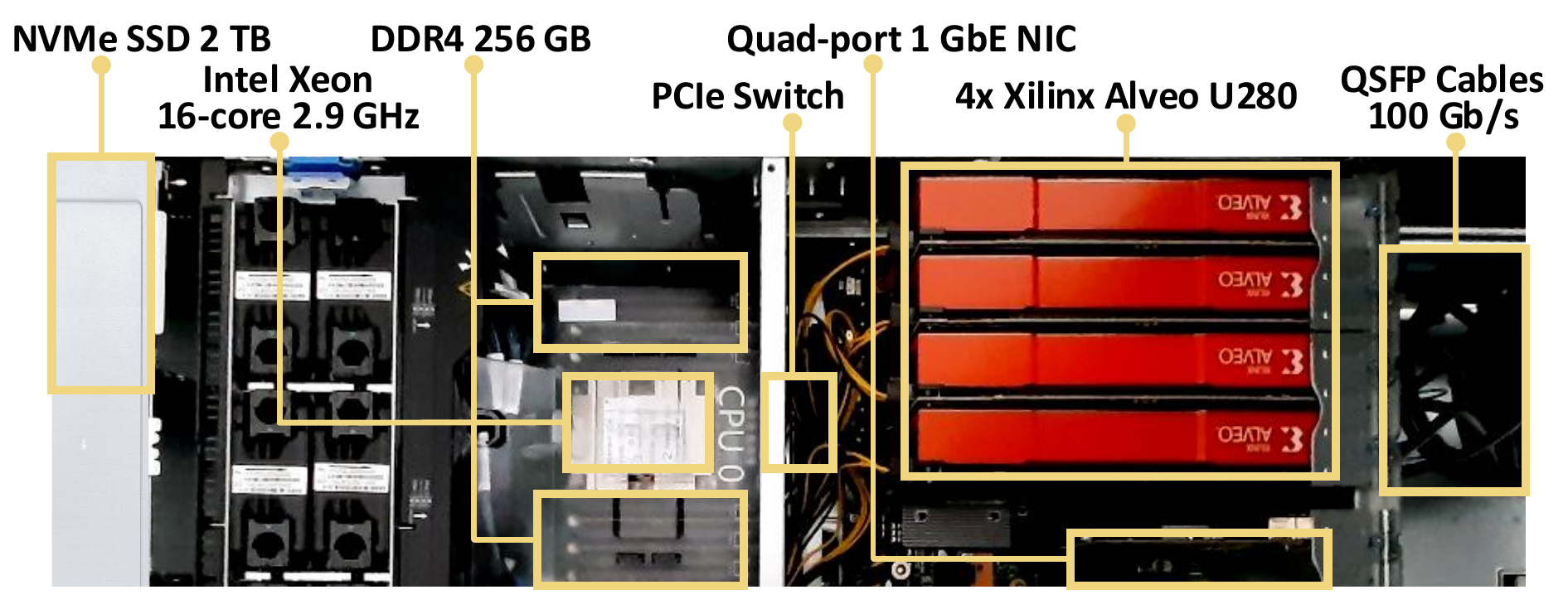}
\vspace{-0.15in}
\caption{Image of \sysname{} server appliance.}
\label{fig-server}
\vspace{-0.1in}
\end{figure}

We build a \sysname{} appliance prototype that uses Intel Xeon Gold 6226R CPU and four Xilinx Alveo U280 data center acceleration cards \cite{alveo} for evaluation. Although it uses a single homogeneous multi-FPGA cluster with four FPGA cards, the appliance itself is capable of harnessing two sets of these configurations and/or increasing the number FPGA cards per cluster in a 4U server chassis. The server appliance can be easily extended as it is already based on a dual-socket motherboard with 20 PCIe Gen3 x16 slots. Installing more FPGA cards and configuring their clusters at the system would be sufficient. Figure \ref{fig-server} shows the setup of the \sysname{} server appliance hardware.

%%%%%%%%%%%%%%%%%%%%%%%%%%%%%%%%%%%%%%%%%%%%%%%%%%
%%%%%%%%%%%%%%%%%%%%%%%%%%%%%%%%%%%%%%%%%%%%%%%%%%

%%%%%%%%%%%%%%%%%%%%%%%%%%%%%%%%%%%%%%%%%%%%%%%%%%
\begin{figure}[t]
\centering
\includegraphics[width=3.33in]{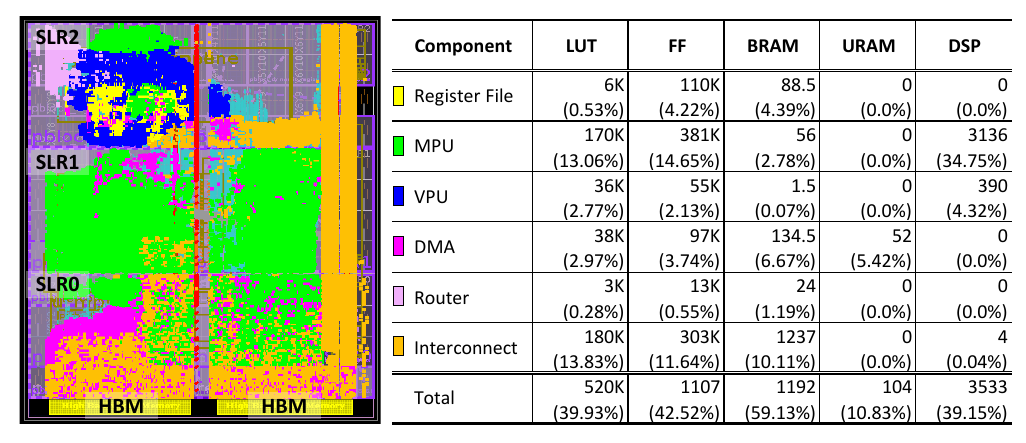}
\vspace{-0.15in}
\caption{FPGA layout and resource utilization on Xilinx Alveo U280.}
\label{fig-fpga}
\vspace{0.05in}
\end{figure}
%%%%%%%%%%%%%%%%%%%%%%%%%%%%%%%%%%%%%%%%%%%%%%%%%%

%%%%%%%%%%%%%%%%%%%%%%%%%%%%%%%%%%%%%%%%%%%%%%%%%%
\begin{figure*}[t]
%\vspace{0.01in}
\centering
\includegraphics[width=6.95in]{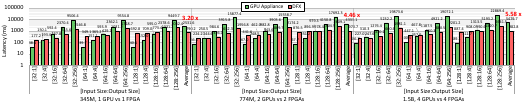}
\vspace{-0.1in}
\caption{Inference latency of \sysname{} compared to the GPU appliance on various GPT-2 models.}
\label{fig-latency}
% \vspace{-0.02in}
\end{figure*}
%%%%%%%%%%%%%%%%%%%%%%%%%%%%%%%%%%%%%%%%%%%%%%%%%%

U280 FPGA chip uses a chiplet-based (i.e., multi-die) design with three super logic regions (SLRs) and supports 8 GB of HBM with 32 channels. We integrate 4 \sysname{} cores, 1 core per device, across the four FPGAs. 
%Since the target application is memory-bound, our design focus is to utilize all the available 32 channels of HBM to feed the massively-parallel matrix unit. 
Considering the total bitwidth of data path is 32$\times$512, the place-and-route (PnR) over three different dies is challenging. Since the HBM controller is physically located in the bottom SLR (i.e., SLR0) and the \sysname{} core is large enough over the dies, we discover that the implementation is eventually constrained by the number of super long line routes (SLLs) which connect the logic between the dies.
%The \sysname{} core is large enough to require hardware resources in all three SLRs, but the implementation of a core that spans across the SLRs is constrained by the number of super long line routes (SLL) that connect the logic die-to-die. 
To handle this multi-die crossing issue effectively, we first decide to split the \sysname{} microarchitecture into kernels, in which the kernel represents a top-level design module to be implemented within a single die.
%an encapsulating top-level module that will be implemented within a single die. 
Then, we map the kernel with the DMA and MPU modules into the SLR closest to the HBM, or SLR0, because these modules frequently access the memory interconnect that requires the most amount of routing; otherwise, these routings would need to be SLR-crossing and exceed the number of available SLLs. However, when we try to map all the MPU lanes and memory channels in a single die, the kernel causes PnR failure due to routing congestion. 
%It is understandable as the interconnect for 32 channels consumes a large amount of hardware resources as well.
%However, an interconnect that utilizes all 32 HBM channel's bandwidth also consume large FPGA resources. Thus, the kernel causes place-and-route failure due to routing congestion. 
As a result, we map the maximum possible lanes of MPU that can meet the routing constraint of SLR0 and map the rest into the other SLRs. 
%Lastly, the result from the matrix-vector multiplication in the MPU is a subvector of only $lane$ dimension, so the register file and VPU are placed in a separate SLR. 
%For the final implementation, we place and route the kernel with the DMA and 6 lanes of MPU into SLR0, the kernel with 10 lanes of MPU into SLR1, and the kernel with the register file, VPU, and router into SLR2.
%The communication between the kernels implemented in different SLR are enabled through kernel to kernel streaming (K2K) that utilizes high-speed and lightweight AXI-Stream interface. 
By having separate kernels that minimize die-crossing signals, the \sysname{} core overcomes the routing congestion and achieves the maximum bandwidth utilization out of the device. 
%\sysname{} core overcomes the routing congestion and timing violation that occur when maximizing logic usage and utilizing high number of memory channels of the HBM.

We run all U280 FPGAs at 200 MHz kernel frequency and 410 MHz memory interface frequency, utilizing 39.93\% of LUT, 42.52\% of FF, 59.13\% of BRAM, 10.83\% of URAM, and 39.15\% of DSP. Figure \ref{fig-fpga} shows the final FPGA layout and resource utilization of one of the U280s. We use Xilinx Vivado \cite{vivado} to synthesize the hardware written in SystemVerilog and the Xilinx Vitis 2020.2 platform \cite{vitis} for the host-FPGA communication.

%%%%%%%%%%%%%%%%%%%%%%%%%%%%%%%%%%%%%%%%%%%%%%%%%%%%%%%%%%%%%%%%%%%%%%%%%
\begin{table}[t]
% \vspace{-0.16in}
\centering
\scriptsize
\caption {GPT-2 Model Configuration} \label{tab:config} 
\vspace{-0.1in}

\begin{tabular}{c|c|c|c|c} 
\toprule
\begin{tabular}[c]{@{}c@{}}\textbf{Number of}\\\textbf{Parameters}\end{tabular} & 
\begin{tabular}[c]{@{}c@{}}\textbf{Embedding}\\\textbf{Dimension}\end{tabular} & 
\begin{tabular}[c]{@{}c@{}}\textbf{Number of}\\\textbf{Attention Heads}\end{tabular} & 
\begin{tabular}[c]{@{}c@{}}\textbf{Head}\\\textbf{Dimension}\end{tabular} & 
\begin{tabular}[c]{@{}c@{}}\textbf{Number of}\\\textbf{Layers}\end{tabular}  \\ 
\hline\hline
345M & 1024 & 16 & 64 & 24 \\ 
\hline
774M & 1280 & 20 & 64 &  36 \\ 
\hline
1.5B & 1536 & 24 & 64 &  48 \\
\bottomrule
\end{tabular}
\vspace{0.1in}
\end{table}

%%%%%%%%%%%%%%%%%%%%%%%%%%%%%%%%%%%%%%%%%%%%%%%%%%%%%%%%%%%%%%%%%%%%%%%%%

% \vspace{-0.05in}

\section{Evaluation}
\label{evaluation}

We use the \sysname{} appliance prototype to evaluate the system performance. We use a GPU appliance, a custom server of four NVIDIA V100 GPUs \cite{v100}, as the evaluation baseline to compare the results with \sysname{}.
% \textcolor{red}{For a fair comparison, we choose V100 GPU, which is as similar as possible to U280 FPGA's hardware specification, especially the memory capacity and bandwidth.}  
The V100 GPU has the most comparable hardware specification to the U280 FPGA, especially the memory capacity and bandwidth, to yield a fair comparison. We run the GPT-2 models on this GPU appliance using NVIDIA's GPU-optimized Megatron-LM source code \cite{shoeybi2019megatron} and CUDA Toolkit 11.1 with the provided parallelism scheme that supports both multi-GPU training and inference. In addition, we run the cloud TPU \cite{jouppi2017datacenter, tpucloud} to analyze the performance of different accelerator platforms. Regarding each model, we use the open-source 345M model from NVIDIA Megatron-LM \cite{shoeybi2019megatron}, and 774M and 1.5B model from OpenAI \cite{radford2019language}. We slightly adjust OpenAI’s 1.5B model’s number of attention heads from 25 to 24 because OpenAI’s configuration is difficult to parallelize for both hardware platforms to run. Table \ref{tab:config} shows the GPT-2 configuration for each model. We use Xilinx Board Utility (xbutil)  and NVIDIA system management interface (nvidia-smi) for power measurements. We measure the inference accuracy, latency, throughput, and energy efficiency. Furthermore, we discuss the scalability of \sysname{} and compare the cost of the two systems.

\subsection{Inference Accuracy}

\textbf{Methodology}
To ensure that \sysname{} does not incur any accuracy loss on GPT-2, we compare the accuracy for widely used open-source datasets with the baseline V100 GPU on the 345M model. We compare Winograd Schema Challenge (WSC) \cite{levesque2012winograd}, Childrens' Book Common Noun (CBT-CN), and Children's Book Named Entities (CBT-NE) \cite{bajgar2016embracing}, which predict a word based on the given context.

\textbf{Accuracy}
% Table \ref{tab:accuracy} shows the accuracy comparison between \sysname{} and the GPU appliance with NVIDIA V100 GPUs. 
\sysname{} achieves no loss, 0.3\% loss, and 0.15\% gain in accuracy for WSC, CBT-CN, and CBT-NE, respectively, when compared to the baseline GPU. The GPU runs its kernel with FP16 operators, a standard for NLP applications. To minimize the error in text generation applications, \sysname{} also runs its cores with FP16 operators based on the Xilinx Floating-Point Operator IP. Both FP16 operators are based on IEEE 754 with 1-bit sign, 5-bit exponent, and 10-bit mantissa. Since all the operations are identical in both systems except for the GELU operation, the difference comes from the subtle difference in approximation between the GPU and \sysname{}, which is negligible.

% %%%%%%%%%%%%%%%%%%%%%%%%%%%%%%%%%%%%%%%%%%%%%%%%%%%%%%%%%%%%%%%%%%%%%%%%%
% \begin{table}[t]
% \vspace{0.01in}
% \centering
% \scriptsize
% \caption {GPT-2 Accuracy results} \label{tab:accuracy} 

% \begin{tabular}{c|c|c|c} 
% \toprule
%     & \textbf{WSC} & \textbf{CBT-CN} & \textbf{CBT-NE}  \\ 
% \hline\hline
% GPU Appliance   & 59.65\%   & 89.50\%   & 79.15\% \\ 
% \hline
% DFX             & 59.65\%   & 89.20\%   & 79.30\% \\
% \hline
% \hline
% Accuracy Loss   & 0.0\%   & 0.3\%    & -0.15\%  \\
% \bottomrule
% \end{tabular}
% \vspace{-0.1in}
% \end{table}

% %GPU Appliance  & 58.02\%  & 87.28\%    & 69.00\%          \\ 
% %DFX            & 60.35\%  & 86.26\%    & 67.00\%          \\
% %Accuracy Loss  & -2.33\%  & 1.02\%     & 2.00\%          \\

% %%%%%%%%%%%%%%%%%%%%%%%%%%%%%%%%%%%%%%%%%%%%%%%%%%%%%%%%%%%%%%%%%%%%%%%%%

% \vspace{-0.015in}

\subsection{Performance Analysis}
\label{performance}

\textbf{Methodology}
We evaluate the end-to-end performance of \sysname{} on various GPT-2 models (345M, 774M, and 1.5B) with different combinations of input and output token lengths to represent the dialogue system, topic-to-essay generation, and other text generation workloads. We use the same model and the same number of accelerators in both appliances for a fair comparison.
%We run the model on \sysname{} and compare it against the GPU appliance.
We compare one V100 GPU and one U280 FPGA for the 345M model, two V100 GPUs and two U280 FPGAs for the  774M model, and four V100 GPUs and four U280 FPGAs for the 1.5B model. Specifically, we run each model with input lengths of 32, 64, and 128 tokens and various output lengths between 1 and 256 tokens, which are the typical ranges of user requests for transformer-based language services in datacenters\cite{tokenratio}. 

% and expand the model in-house to create the 2.5B and 8.3B models.

\textbf{Latency} 
Figure \ref{fig-latency} shows the text generation latency of the GPU appliance and \sysname{} on various GPT-2 models, as well as the speedup that \sysname{} achieves. Note that the Y-axis for latency is in log scale. The result shows that \sysname{} achieves an average of 5.58$\times$ speedup compared to the GPU appliance for the 1.5B model. For the workload with significantly more output tokens compared to the input token, i.e., 32:256, \sysname{} shows substantially lower latency, 10.03$\times$, than the GPU appliance. For the 345M and 774M models, \sysname{} achieves an average of 3.20$\times$ and 4.46$\times$ speedup, respectively, compared to the GPU appliance with the equivalent number of accelerators. The result presents that the additional number of output tokens leads to a more significant increase in latency on the GPU appliance than on \sysname{}. In fact, the speedup of \sysname{} over the GPU appliance can be greater for even smaller input and larger output sizes. Although the future NLG applications may require longer token generations, we only focus on the range of current use cases. 
%It is noteworthy that \sysname{} is adaptable to any configuration with a programmable core tailored for text generation.
The overall speedup of \sysname{} is attenuated with a larger input size because the GPU is able to take advantage of its massively parallel computation to execute the large input. As long as the ratio between the input and output lengths is lower than 4:1, which is the case for text generation workloads, \sysname{} performs better than the GPU appliance. 

%%%%%%%%%%%%%%%%%%%%%%%%%%%%%%%%%%%%%%%%%%%%%%%%%%
\begin{figure}[t]
% \vspace{-0.08in}
\centering
\includegraphics[width=2.9in]{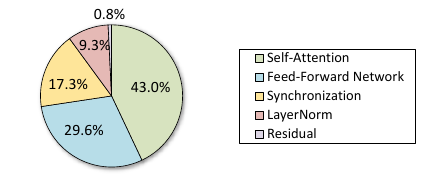}
\vspace{-0.1in}
\caption{Latency breakdown of 4 FPGAs on the 1.5B model.}
\label{fig-breakdown2}
% \vspace{0.05in}
\end{figure}
%%%%%%%%%%%%%%%%%%%%%%%%%%%%%%%%%%%%%%%%%%%%%%%%%%

%%%%%%%%%%%%%%%%%%%%%%%%%%%%%%%%%%%%%%%%%%%%%%%%%%
\begin{figure}[t]
% \vspace{0.01in}
\centering
\includegraphics[width=2.9in]{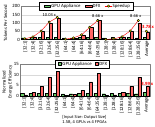}
\vspace{-0.1in}
\caption{Throughput and energy efficiency of \sysname{} compared to the GPU appliance on the 1.5B model.}
\label{fig-throughput}
\vspace{0.1in}
\end{figure}
%%%%%%%%%%%%%%%%%%%%%%%%%%%%%%%%%%%%%%%%%%%%%%%%%%

%%%%%%%%%%%%%%%%%%%%%%%%%%%%%%%%%%%%%%%%%%%%%%%%%%
\begin{figure}[t]
% \vspace{0.05in}
\centering
\includegraphics[width=2.8in]{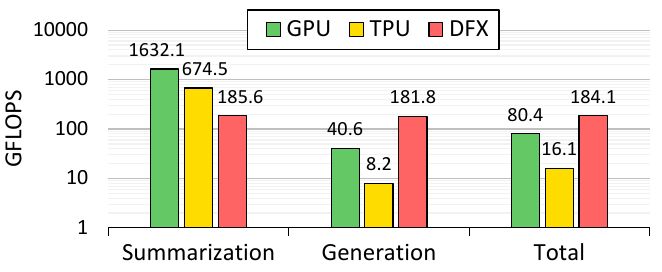}
\vspace{-0.1in}
\caption{Performance comparison with GPU, TPU, and DFX (1 FPGA) on the 345M model.}
\label{fig-gops}
% \vspace{-0.1in}
\end{figure}
%%%%%%%%%%%%%%%%%%%%%%%%%%%%%%%%%%%%%%%%%%%%%%%%%%

%%%%%%%%%%%%%%%%%%%%%%%%%%%%%%%%%%%%%%%%%%%%%%%%%%
\begin{figure}[t]
\centering
\includegraphics[width=2.2in]{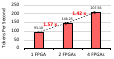}
\vspace{-0.1in}
\caption{Scalability of \sysname{} on the 345M model.}
\label{fig-scalability}
\vspace{0.1in}
\end{figure}
%%%%%%%%%%%%%%%%%%%%%%%%%%%%%%%%%%%%%%%%%%%%%%%%%%

Figure \ref{fig-breakdown2} shows the latency breakdown of running the 1.5B model on 4 FPGAs. The result shows that the majority of the time is consumed by self-attention and feed-forward network at 72.6\%. At 17.3\%, synchronization may seem critical in \sysname{} when compared to the GPU appliance because GPU has high-speed communication such as NVLink \cite{li2019evaluating} to lower the synchronization latency. However, \sysname{} has 5.58$\times$ lower overall latency, so the high synchornization proportion is attributed to the speedup in other operations. %This claim is supported by the fact that synchronization only happens four times per layer, as mentioned in Section \ref{architecture_cluster}.

%Although the synchronization overhead between GPUs is negligible due to its high-speed communication via NVLink \cite{li2019evaluating}, GPU performance does not scale because it is limited by its compute units that underperform for the sequential process in text generation. Hence, adding more GPU devices results in more underutilization without any performance gain, which confirms that GPU-based platform less suitable for text generation workloads.

%Meanwhile, both \sysname{} and the GPU appliance effectively support models of various scales. 
% Specifically, the \sysname{} architecture efficiently maps large models as shown in Figure \ref{fig-breakdown2}, in which the percentage of time spent on synchronization does not increase as the model size increases.

\textbf{Throughput and Energy Efficiency} 
Figure \ref{fig-throughput} shows the throughput and energy efficiency of the two appliances on the 1.5B model. \sysname{} achieves an average of 3.78$\times$ throughput and 3.99$\times$ high energy efficiency compared to the GPU appliance. The throughput was measured by dividing the number of output tokens by the text generation latency. The throughput result shows that GPU maintains a relatively constant throughput even when the output tokens are scaled up, which indicates that the performance is bottlenecked by low hardware utilization during the generation stage. Furthermore, we observe that 1) each V100 GPU consumes only 47.5W, on average, based on the nvidia-smi tool, and 2) the average power consumption decreases as the number of output tokens increases. Since the GPU runs at a high base clock frequency of 1.23GHz, the low power consumption can only be explained by low hardware utilization. Meanwhile, \sysname{} runs at an even lower 45W, not because of low hardware utilization but because the FPGA runs at a lower operating frequency of 200MHz.

%Measure sumamrization and generation stage power separately

We also analyze the GFLOPS performance of different accelerator platforms, DFX accelerator (1 FPGA), TPU, and GPU, for the 345M model with 64:64 tokens, as shown in Figure \ref{fig-gops}. The GPU and TPU show similar behavior with high throughput in the summarization stage (1632.1 and 674.5 GFLOPS) and signficantly reduced throughput in the generation stage (40.6 and 8.2 GFLOPS), which implies that these devices are highly utilized in a batched process but severely underutilized in a non-batchable process. Meanwhile, DFX retains an average of 184.1 GFLOPS during both summarization and generation stages because its dataflow is specialized for the iterative matrix-vector multiplication rather than the infrequent matrix-matrix multiplication.

\textbf{Scalability}
Figure \ref{fig-scalability} shows the scalability of U280 FPGA in \sysname{} for the 345M model with 64:64 tokens. On average, \sysname{} achieves 93.10 tokens/sec for 1 FPGA, 146.25 tokens/sec for 2 FPGAs, and 207.56 tokens/sec for 4 FPGAs. The performance of \sysname{} increases linearly with the number of FPGAs at the rate of 1.5, which means the processing unit for \sysname{} is designed to retain high utilization with more devices, even on relatively small models. The performance gain is not directly proportional to the number of devices because we do not parallelize layer normalization and residual due to their even larger synchronization overhead. There is also a marginal drop in throughput with each additional FPGA due to more data synchronization. 

%Although the synchronization overhead between GPUs is negligible due to its high-speed communication via NVLink \cite{li2019evaluating}, GPU performance does not scale because it is limited by its compute units that underperform for the sequential process in text generation. Hence, adding more GPU devices results in more underutilization without any performance gain, which confirms that GPU-based platform less suitable for text generation workloads.

%%%%%%%%%%%%%%%%%%%%%%%%%%%%%%%%%%%%%%%%%%%%%%%%%%%%%%%%%%%%%%%%%%%%%%%%%
\begin{table}[t]
% \vspace{-0.01in}
\centering
\scriptsize
\caption {Appliance Cost Analysis} \label{tab:cost} 
\vspace{-0.1in}

\begin{tabular}{l|l|l} 
\toprule
              & \textbf{GPU Appliance} & \textbf{DFX}                                                               \\ 
\hline\hline
CPUs          &   \begin{tabular}[c]{@{}l@{}}2 $\times$ Intel Xeon Gold\\14-Core @2.2 GHz\end{tabular}      & \begin{tabular}[c]{@{}l@{}}2 $\times$ Intel Xeon Gold\\16-Core @2.9 GHz\end{tabular}  \\
\hline
Memory        &  384 GB DDR4                                                                     & 512 GB DDR4                                                               \\ 
\hline
Storage       & 12 TB NVMe                                                                       & 4 TB NVMe                                                                          \\ 
\hline
Accelerators  &  \begin{tabular}[c]{@{}l@{}} 4 $\times$ NVIDIA Tesla V100\\32 GB HBM2 (900 GB/s)\end{tabular}     & \begin{tabular}[c]{@{}l@{}} 4 $\times$ XILINX Alveo U280\\8 GB HBM2 (460 GB/s)\end{tabular} \\ 
\hline
Performance     & 13.01 tokens/sec    & 72.68 tokens/sec                                                                       \\                                                      
\hline
Cost           & \$45,832 (\$11,458 per GPU) &   \$31,180 (\$7,795 per FPGA) \\
\hline\hline
\begin{tabular}[c]{@{}l@{}}Performance\\/ Cost\end{tabular}  &    283.86 tokens/sec/million\$   & 2330.98 tokens/sec/million\$ \\
\bottomrule
\end{tabular}
\vspace{0.1in}
\end{table}
%%%%%%%%%%%%%%%%%%%%%%%%%%%%%%%%%%%%%%%%%%%%%%%%%%%%%%%%%%%%%%%%%%%%%%%%%

\subsection{Cost Analysis}
Table \ref{tab:cost} shows the cost analysis of \sysname{} and the GPU appliance. \sysname{} has an upfront cost that is \$14,652 lower than that of the GPU appliance when 4 devices are installed on both appliances \cite{v100cost, v100cost-a, u280cost}. For comparison, we exclude the cost of components (e.g., CPU and storage) other than the accelerators. To measure the overall cost-effectiveness, we consider both performance and upfront cost (i.e., retail price). We choose the 1.5B model with the input-to-output token ratio of 64:64 to measure the performance per cost, which is representative of the chatbot service as described in Section \ref{background_gpt}. 
% because this configuration is most commonly used in datacenters. 
Comparing the performance-to-cost ratio, \sysname{} is 8.21$\times$ more cost-effective than the GPU appliance. %Note that the cost-effectiveness goes up to A$\times$ and B$\times$ if the input-to-output token ratio increases to C and D, respectively.

We can draw similar conclusions between the \sysname{} and the NVIDIA DGX system because our custom GPU appliance is comparable in cost and performance to DGX-1 \cite{dgxcost}. DGX-1 is a purchasable GPU-based appliance that harnesses 8 Tesla V100 GPUs for machine learning workloads in datacenters. Our custom GPU appliance is an upgraded version of DGX-1 with the same number of GPUs, in which each GPU has a better performance grade and a larger HBM2 than those of DGX-1. As the GPU-based appliance would perform worse when scaled up to 8 devices, the stated improvement in cost-effectiveness for \sysname{} would be even greater when compared to real datacenter appliances.

% \vspace{0.1in}

% \vspace{-0.1in}

\section{Related Work}
\label{related_work}

%\vspace{-0.025in}

\textbf{Accelerator for Transformer Model}
Hardware accelerators that support transformer-based NLP models have been recently proposed. A3 \cite{ham2020a3}, GOBO \cite{zadeh2020gobo}, SpAtten \cite{wang2021spatten}, EdgeBERT \cite{tambe2021edgebert}, and ELSA \cite{ham2021elsa} discuss designs that only speed up the attention mechanism in the transformer using various pruning, quantization, or both, without taking into consideration the end-to-end process. For instance, SpAtten uses pruning to accelerate the attention mechanism and the layer normalization process but does not consider token embedding, residual, and LM head. Our work targets the language services at the datacenter, especially focusing on text generation workloads, so there is less emphasis on aggressively speeding up a specific operation in the transformer but more emphasis on the model-parallel dataflow that can address both memory and compute-intensive problems in its end-to-end inference.

\textbf{Hardware Architecture for Datacenters} Many domain-specific acceleration architectures have been proposed for machine learning but few at the scale of datacenters. Microsoft's Brainwave \cite{fowers2018configurable} is a relevant work that implements an FPGA-based neural processing unit for datacenters with high-performance processing units. However, Brainwave does not utilize high-bandwidth off-chip memory, so it cannot effectively run memory-intensive workloads, such as GPT inference. Google \cite{jouppi2017datacenter, jouppi2020domain, jouppi2021ten} proposes the systolic-based TPU architecture for various DNN applications, and Facebook \cite{anderson2021first} designs accelerators specialized for recommendation systems. A few of these previous architectures for datacenters \cite{fowers2018configurable, jouppi2021ten} support the transformer, but none have design optimizations to drastically improve the performance for text generation workloads specifically.

%None of these previous architectures for datacenters support the transformer because the transformer has processes like attention that have different operational characteristics than other DNNs.

% \vspace{0.1in}

% \vspace{-0.1in}

\section{Conclusion}
\label{conclusion}

%\vspace{-0.03in}

This paper presents \sysname{}, a low-latency multi-FPGA appliance for accelerating transformer-based text generation workloads. Our work identifies the need for designing a datacenter-level system that provides low-latency inference, end-to-end acceleration, and parallel computing for text generation. We combine multi-device hardware with model parallelism, custom instructions and dataflow, and other hardware optimizations to maximize the potential of the specified hardware. Based on the implementation results of our DFX appliance prototype with four Xilinx Alveo U280 FPGAs, \sysname{} achieves 5.58$\times$, 3.99$\times$, and 8.21$\times$ improvements in performance, energy-efficiency, and cost-effectiveness, respectively, compared to conventional multi-GPU appliances. 

% The implementation results of our proposed architecture on four Xilinx Alveo U280 FPGAs show that \sysname{} ... "
% \vspace{0.1in}

\section*{Acknowledgements}

This research was supported in part by NAVER CLOVA and the MSIT (Ministry of Science and ICT), Korea, under the ITRC (Information Technology Research Center) support program (IITP-2020-0-01847) supervised by the IITP (Institute for Information \& Communications Technology Planning \& Evaluation). 
%This research was funded by NAVER CLOVA. The experiments were mostly conducted on the CastLab FPGA platform and partly conducted on the GPU server of NAVER CLOVA.

%%%%%%% -- PAPER CONTENT ENDS -- %%%%%%%%

% trigger a \newpage just before the given reference
% number - used to balance the columns on the last page
% adjust value as needed - may need to be readjusted if
% the document is modified later
%\IEEEtriggeratref{8}
% The "triggered" command can be changed if desired:
%\IEEEtriggercmd{\enlargethispage{-5in}}

% references section

% can use a bibliography generated by BibTeX as a .bbl file
% BibTeX documentation can be easily obtained at:
% http://www.ctan.org/tex-archive/biblio/bibtex/contrib/doc/
% The IEEEtran BibTeX style support page is at:
% http://www.michaelshell.org/tex/ieeetran/bibtex/
%\bibliographystyle{IEEEtran}
% argument is your BibTeX string definitions and bibliography database(s)
%\bibliography{IEEEabrv,../bib/paper}
%
% <OR> manually copy in the resultant .bbl file
% set second argument of \begin to the number of references
% (used to reserve space for the reference number labels box)
% \begin{thebibliography}{1}

% \bibitem{IEEEhowto:kopka}
% H.~Kopka and P.~W. Daly, \emph{A Guide to \LaTeX}, 3rd~ed.\hskip 1em plus
%   0.5em minus 0.4em\relax Harlow, England: Addison-Wesley, 1999.

% \end{thebibliography}

\bibliographystyle{IEEEtran}
\bibliography{refs}

% that's all folks
\end{document}